\documentstyle[11pt,aasms4,epsf]{article}

\def\folio{\ifnum\pageno<2\nopagenumbers\else\number\pageno\fi}
\newtoks\headline \headline={\hss\twelverm\folio\hss} 
\newtoks\footline \footline={{\hfil}} 
\font\mathbf=cmmib10 scaled 1000             

\def\ref{\par\noindent\hangindent=2pc \hangafter=1 }
\def\amin{\ifmmode^{\prime}\else$^{\prime}$\fi}
\def\asec{\ifmmode^{\prime\prime}\else$^{\prime\prime}$\fi}
\def\dasec  {\hbox{$.\!\!^{\prime\prime}$}}     

\def\etal{{et al. }}
\def\cappage #1 #2 #3 {\vfill\eject\pageno=#1
\vglue 10 true in minus 10 true in \noindent{\bf Figure #2.} #3}
\def\ee #1 {\times 10^{#1}}
\def\ut #1 #2 { \, \hbox{#1}^{#2}}
\def\u #1 { \, \hbox{#1}}

\def\msol{\, \hbox{$\hbox{M}_\odot$}}

\let\grad=\nabla
\def\cross{{\bf \times}}
\def\curl #1 {\grad \cross #1}
\def\div #1 {\grad \cdot #1}

\def\msol   {\hbox{$M_\odot$}}                  
\def\etal   {{\it et al. }}                     

\begin{document}

\title{A Multi-Wavelength  Study of Sgr A*:  
The Role of Near-IR Flares in Production of X-ray, 
Soft $\gamma$-ray and  Submillimeter Emission}

\author{F. Yusef-Zadeh\footnote{Department of Physics and Astronomy, 
Northwestern University, Evanston, Il. 60208 
(zadeh@northwestern.edu)},
 H. Bushouse\footnote{STScI, 3700 San Martin Drive, 
Baltimore, MD  21218 (bushouse@stsci.edu)}, 
C.D. Dowell\footnote{Cal Tech, Jet 
Propulsion Laboratory, Pasadena, CA 91109 
(cdd@submm.caltech.edu)}, 
M. Wardle\footnote{
Department of Physics, Macquarie University, Sydney NSW 2109, 
Australia (wardle@physics.mq.edu.au)}, 
D. Roberts\footnote{Adler Planetarium and Astronomy Museum, 1300 
South Lake Shore Drive, Chicago, IL 60605
(doug-roberts@northwestern.edu)}, 
C. Heinke
\footnote{Department of Physics and 
Astronomy,
Northwestern University, Evanston, Il. 60208 
(cheinke@northwestern.edu)}, 
G. C. Bower\footnote{Radio Astronomy Lab, 601 Campbell Hall, 
University of California, Berkeley, CA 94720
(gbower@astron.berkeley.edu)}
B. Vila Vilaro \footnote{ 
National Observatory of Japan
2-21-1 Osawa, Mitaka, Tokyo 181-8588, Japan
(vila.vilaro@nao.ac.jp)}, 
S. Shapiro\footnote{Department of Physics, University of 
Illinois at Urbana-Champaign, Urbana, IL 61801-3080 
(shapiro@astro.physics.uiuc.edu)}, 
A. Goldwurm\footnote{Service d'Astrophysique / CEA-Saclay, 91191 Gif sur Yvette, France;
Astroparticle and Cosmology, 11 place Berthelot, 75005, Paris, France
(goldwurm@discovery.saclay.cea.fr)}
G. Belanger\footnote{Service d'Astrophysique / CEA-Saclay, 91191 Gif sur Yvette, France;
Astroparticle and Cosmology, 11 place Berthelot, 75005, 
Paris, France(belanger@cea.fr)}}
 
\begin{abstract}

Although Sgr A* is known to be variable in radio, millimeter, near-IR and 
X-rays, the correlation of the variability across its spectrum has not 
been fully studied. Here we describe highlights of the results of two 
observing campaigns in 2004 to investigate the correlation of flare 
activity in different wavelength regimes, using a total of nine ground and 
space-based telescopes. 
We report the detection of  several new near-IR flares during the campaign 
based on 
{\it HST} observations. The 
level of near-IR flare activity 
can be as low as  $\sim0.15$ mJy at 1.6 $\mu$m 
and continuous  up to 
$\sim$40\% of the total observing  time, thus placing better limits than 
ground-based 
near-IR observations. 
Using the 
NICMOS instrument on the {\it HST}, the {\it 
XMM-Newton} and {\it Caltech Submillimeter} observatories, 
we also detect 
simultaneous bright X-ray 
and near-IR flare in which we observe for the first time 
correlated substructures 
as well as  simultaneous  submillimeter and 
near-IR flaring. 
X-ray emission is  arising from the population of 
near-IR-synchrotron-emitting relativistic particles which  scatter
submillimeter seed photons within the inner 10 Schwarzschild radii 
(R$_{sch}$) of Sgr A* up to X-ray energies. In addition, using the inverse 
Compton scattering picture, we explain the high energy 20-120 keV emission 
from the direction toward Sgr A*, and the lack of one-to-one X-ray 
counterparts to near-IR flares, by the variation of the magnetic field and 
the spectral index distributions of this population of nonthermal 
particles. In this picture, the evidence for the variability of 
submillimeter emission during a near-IR flare is produced by the 
low-energy component of the population of particles emitting synchrotron 
near-IR emission.  Based on the measurements of the duration of flares 
in near-IR and submillimeter wavelengths, we argue that the
cooling could  be  due to  adiabatic expansion with the implication 
that  flare activity may drive an outflow. 


\end{abstract}



\section{Introduction}

More than three decades have elapsed since the discovery of Sgr 
A* (Balick \& Brown 1974) and during most of this time the 
source
remained undetected outside the radio band.
Submillimeter radio emission (the ``submillimeter bump'') and 
both
 flaring and quiescent
X-ray emission from Sgr A* are now believed to originate within 
just
a few 
Schwarzschild radii of the $\sim3.7\times10^6$ \msol\ 
black hole
(Baganoff et al.\ 2001; 
Sch\"odel et al. 2002; Porquet et al. 2003; Goldwurm et al. 2003;
Ghez et al. 2005).
Unlike the most powerful X-ray flares which  show    a soft  spectral 
index
(Porquet et al. 2003), most X-ray flares from Sgr A*  are weaker and have hard 
spectral indices.
More recently, the
long-sought near-IR counterpart to Sgr A$^*$ was discovered
 (Genzel et al. 2003). During
several near-IR flares (lasting  $\sim$40 minutes) Sgr A*'s flux
increased by a factor of a
few (Genzel et al. 2003; Ghez et
al. 2004). Variability has also been seen at centimeter and
millimeter wavelengths with a time scale ranging between hours to 
years with amplitude variation at a level of  less than 100\% (Bower et 
al. 2002; 
Zhao et 
al.\ 2003; Herrnstein et al.\ 2004; Miyazaki et al. 2004; Mauerhan et al.
2005). These variations are  at much lower level than observed at
 near-IR and X-ray  wavelengths.
Recently, Macquart \& Bower (2005) have shown that the radio
and millimeter flux density variability on time scales longer than a few
days can be explained through interstellar scintillation.

Although the discovery of bright X-ray flares from Sgr A* has 
helped us to understand how mass accretes onto black holes at 
low accretion rates, it has left many other questions 
unanswered.  The simultaneous observation of Sgr A* from radio 
to $\gamma$-ray can be helpful for distinguishing among the 
various emission models for Sgr A* in its quiescent phase and 
understanding the long-standing puzzle of the 
extremely low accretion rate deduced for Sgr A*. Past 
simultaneous observations to measure the correlation of the 
variability over different wavelength regimes have been 
extremely limited. Recent work by Eckart et al. (2004, 2005) 
detected 
 near-IR counterparts to the decaying part of an X-ray flare as 
well as a full X-ray  flare based on Chandra observations. 

In order to 
obtain a more complete wavelength coverage across its spectrum, 
Sgr~A$^*$ was the focus of an organized and unique observing 
campaign at radio, millimeter, submillimeter, near-IR, X-ray 
and soft $\gamma$-ray wavelengths. This campaign was intended to 
determine the physical mechanisms responsible for accretion 
processes onto compact objects with extremely low luminosities 
via studying the variability of Sgr~A*.  The luminosity of 
Sgr~A* in each band is known to be about ten  orders of magnitudes 
lower than the Eddington luminosity, prompting a number of 
theoretical models to explain its faint quiescent as well as its 
flaring X-ray and near-IR emission in terms of 
the inverse Compton scattering (ICS) of  submillimeter photons
close to the event horizon of Sgr A* (Liu \& Melia 2002; Melia 
and Falcke 2001; Yuan, Quataert \& Narayan  2004; Goldston, 
Quataert \& Tgumenshchev 
2005; Atoyan \& Dermer 2004; Liu, Petrosian,  \&  
Melia 2004;  Eckart et al. (2004, 2005); Markoff 2005).

The campaign consisted of two epochs of observations starting 
March 28, 2004 and 154 days later on August 31, 2004. The 
observations with various telescopes lasted for about four days 
in each epoch. The first epoch employed the following 
observatories: XMM-Newton, INTEGRAL, Very Large Array (VLA) of 
the National Radio Astronomy Observatory\footnote{The National 
Radio Astronomy Observatory is a facility of the National 
Science Foundation, operated under a cooperative agreement by 
Associated Universities, Inc.}, Caltech Submillimeter 
Observatory (CSO), Submillimeter Telescope (SMT), Nobeyama 
Array (NMA), Berkeley Illinois Maryland Array (BIMA) and 
Australian Telescope Compact Array (ATCA).  The second epoch 
observations used only five observatories: XMM-Newton, INTEGRAL, 
VLA, Hubble Space Telescope (HST) Near Infrared Camera and 
Multi-Object Spectrometer (NICMOS) and CSO.  Figure 1 shows 
a schematic diagram showing all of the 
instruments that were used during the two observing campaigns. 
A more detailed account of radio  data will 
be presented elsewhere (Roberts et al. 2005)

An outburst from an 
eclipsing binary CXOGCJ174540.0-290031 took place prior to 
the first epoch and consequently confused the variability 
analysis of Sgr A*, especially in low-resolution data (Bower et al. 2005; 
Muno et al. 2005; Porquet et al. 2005). Thus, 
most of the analysis presented here concentrates on our second 
epoch observations. In addition, ground-based near-IR 
observations of Sgr~A* using the VLT were corrupted in both 
campaigns due to bad weather. Thus, the only near-IR data was 
taken using NICMOS of HST in the second epoch. The structure of 
this paper is as follows. We first concentrate on the highlights of 
variability results of Sgr A* in different wavelength regimes in 
an increasing order of wavelength, followed by the correlation 
of the light curves, the power spectrum analysis of the light curves in 
near-IR wavelengths and construction of its
multiwavelength spectrum.  
We then 
discuss the emission mechanism responsible for the flare 
activity of Sgr A*.

\section{Observations}
\subsection {X-ray and $\gamma$-ray Wavelengths: XMM-Newton and 
INTEGRAL}

One of us (A.G.) was the principal investigator who was 
granted observing time using the  XMM-Newton and INTEGRAL 
observatories to monitor 
the spectral and temporal properties of Sgr A*. These high-energy observations led
the way for other simultaneous observations. Clearly, X-ray 
and 
$\gamma$-ray observations had the most complete time coverage during 
the campaign. A total of 550 ks observing time or 
$\approx$1 week was given to XMM observations, two orbits  
(about 138 ks each) in each of two epochs (Belanger et al. 2005a; 
Proquet et al. 2005). Briefly, these X-ray observations discovered 
two relatively strong flares equivalent of 35 times the quiescent 
X-ray flux of Sgr~A* in each of the two epochs, with peak 
X-ray fluxes of 6.5 and $6\times10^{-12}$ ergs s$^{-1}$ cm$^{-2}$ between 
2-10 keV. These fluxes correspond to X-ray luminosity of 7.6 and 7.7 
$\times10^{34}$ ergs s$^{-1}$ at the distance of 8 kpc, respectively. The 
duration of these flares were about 2500 and 5000 s. 
In addition, the eclipsing X-ray binary system 
CXOGC174540.0-290031 localized within 3$''$ of Sgr~A* was
also detected in both epochs (Porquet et al. 2005). 
Initially, 
the X-ray emission from this transient source was identified  by 
Chandra observation in July 2004 (Muno et al. 2005) before it was realized 
that  its X-ray 
and radio emission persisted during the the first and second epochs of the 
observing campaign (Bower et al. 2005; 
Belanger et al. 2005a; Porquet et al. 2005).
 
Soft $\gamma$-ray observations using INTEGRAL detected a steady 
source IGRJ17456-2901 within $1'$ of Sgr A* between 20-120 keV 
(Belanger et al. 2005b). (Note that the PSF of IBIS/ISGRI of 
INTEGRAL is 13$'$.)   
IGRJ17456-2901 is measured to have a flux 6.2$\times10^{-11}$ erg
s$^{-1}$ cm$^{-2}$ between 20--120 keV corresponding to a luminosity 
of 4.76$\times10^{35}$ erg s$^{-1}$. 
During the time that both X-ray flares occurred, 
 INTEGRAL  missed observing Sgr~A*,  as this  
instrument 
was passing through the radiation belt exactly during these  
X-ray flare events (Belanger et al. 2005b). 

\subsection {Near-IR Wavelengths: HST NICMOS}

\subsubsection {Data Reductions}

As part of the second epoch of the 2004 observing campaign, 32 
orbits of NICMOS observations were granted to study the light curve 
of Sgr~A* in three bands over four days between August 31 and 
September 4, 2004. 
Given that Sgr~A* can be observed for half of each orbit,
the NICMOS observations constituted an excellent near-IR time 
coverage in the second epoch observing campaign. NICMOS camera 1
was used, which has a field of view of $\sim11''$ and a pixel size of 
0.043$''.$ 
Each orbit consisted of two cycles of observations in the broad H-band 
filter (F160W), the narrow-band Pa$\alpha$ filter at 1.87$\mu$m 
(F187N), and an adjacent continuum band at 1.90$\mu$m (F190N). The 
narrow-band F190N line filter was selected to search for 1.87$\mu$m 
line emission 
expected from the combination of gravitational and Doppler effects 
that could potentially shift any line emission outside of the 
bandpass of the F187N. 
Each exposure used the MULTIACCUM readout mode with the predefined
STEP32 sample sequence, resulting in total exposure times of $\sim$7 
minutes per filter with individual readout spacings of 32 seconds.

The IRAF routine ``apphot'' was used to perform aperture photometry
of sources in the NICMOS Sgr~A* field, including Sgr~A* itself.
For stellar sources the measurement aperture was positioned on
each source using an automatic centroiding routine. 
This approach could not be used for measuring Sgr~A*, because its
signal is spatially overlapped by that of the orbiting star S2.
Therefore the photometry aperture for Sgr~A* was positioned by
using a constant offset from the measured location of S2 in each
exposure.
The offset between S2 and Sgr~A* was derived from the orbital
parameters given by Ghez et~al.~(2003).
The position of Sgr~A* was estimated to be 0.13$''$ south 
and 0.03$''$ west  of S2 during the second epoch observing campaign. 
To confirm the accuracy of the position of Sgr~A*,
two exposures of Sgr~A* taken before and during a flare event
were aligned and subtracted, which resulted in an
image showing the location of the flare emission. We believe that earlier 
NICMOS observations  may have not been able to detect 
the variability of Sgr A* due to the closeness of S2 to Sgr A* 
(Stolovy et al. 1999).

At 1.60$\mu$m, the NICMOS camera 1 point-spread function (PSF)
has a full-width at half maximum (FWHM) of ~0.16$''$ or ~3.75 pixels.
Sgr~A* is therefore located at approximately the half-power
point of the star S2.
In order to find an optimal aperature size of Sgr A*, 
excluding signal from S2 which allowed 
enough signal from Sgr A* for a significant detection, several sizes were 
measured. A measurement aperture radius of 2 pixels (diameter of 4 pixels)
was found to be a suitable compromise.

We have made photometric measurements in the F160W (H-band)
images at the 32 second intervals of the individual exposure
readouts.
For the F187N and F190N images, where the
raw signal-to-noise ratio is lower due to the narrow filter
bandwidth, the photometry was performed
on readouts binned to $\sim$3.5 minute intervals.
The standard deviation in the resulting photometry is on the
order of $\sim$0.002 mJy at  F160W (H-band) measurements
and $\sim$0.005 mJy at  F187N and F190N.

The resulting photometric measurements for Sgr~A* show obvious
signs of variability (as discussed below), which we have confirmed 
through comparison with photometry of numerous nearby stars. 
Comparing the light curves of these objects, it is clear that
sources such as S1, S2, and S0-3 are steady emitters, confirming
that the observed variability of Sgr~A* is not due to instrumental
systematics or other effects of the data reduction and analysis.
For example, the light curves of Sgr~A* and star S0-3 in the
F160W band are shown in Figure 2a.
It is clear that the variability of Sgr~A* seen in three of the
six time intervals is not seen for S0-3.
The light curve of IRS 16SW, which is known to be a variable star,
has also been constructed and is clearly consistent with
ground-based observations (Depoy et al. 2004).

\subsubsection {Photometric Light Curves and Flare Statistics}

The thirty-two HST orbits of Sgr~A* observations were distributed in 
six different observing time windows over the 
course of four days of observations. The detected flares are 
generally clustered within three different time windows, as seen
in Figure 2b. 
This figure shows the photometric light curves of Sgr~A* in the
1.60, 1.87, and 1.90$\mu$m NICMOS bands, using a four pixel diameter
measurement aperture.
The observed ``quiescent'' emission level of Sgr~A* in the 1.60$\mu$m band
is $\sim$0.15 mJy (uncorrected for reddening).
During flare events, the emission is seen to increase by
10\%\ to 20\%\ above this level.
In spite of the somewhat lower signal-to-noise ratio for the
narrow-band 1.87 and 1.90$\mu$m data, the flare activity is still
detected in all bands.

Figure 3a presents detailed light curves of Sgr~A* in all 
three NICMOS bands for the three observing time windows that
contained active flare events, which corresponds to the second, 
fourth, and sixth observing windows. 
An  empirical correction has been applied to the fluxes 
in 1.87 and 1.90 $\mu$m bands
in order to overlay them with the 1.60$\mu$m band data.
The appropriate correction factors were derived by computing
the mean fluxes in the three bands during the 
observing windows in which no flares were seen.
This lead us to scale down the observed fluxes in the 1.87 and
1.90$\mu$m bands by factors of 3.27 and 2.92, respectively, in order to
compare the  observed 1.60$\mu$m band fluxes. 
All the data are shown as a time-ordered sequence
in Figure 3a.

Flux variations are detected in all three bands in the 
three observing windows shown in Figure 3a.
The bright flares (top and middle panels) 
show similar spectral and temporal behaviors, both being 
separated by about two days. These bright flares have
multiple components with flux increases of about 20\%\ and 
durations ranging from 2 to 2.5 hours and dereddened 
peak fluxes of $\sim$10.9 mJy at 1.60$\mu$m. The weak 
flares during the end of the fourth observing window (middle panel)
consist of a collection of sub-flares lasting for about 
2--2.5 hours with a flux increase of only 10\%. The light curve 
from the last day of observations, as shown in the bottom panel of  Figure 3a, 
displays the highest level of flare activity over the course 
of the four days. The dereddened peak flux 
at 1.6$\mu$m is $\sim$11.1 mJy and decays in less than 40 minutes. 
Another flare starts about 2 hours later with a rise and fall 
time of about 25   minutes, with a peak dereddened flux of 10.5 mJy 
at 1.6$\mu$m. There are a couple of instances where the flux  
changed  from "quiescent"
level to peak flare level or vice versa in the span of a single
(1 band) exposure, which is on the order of $\sim$7 minutes.
For our 1.6 micron fluxes,
Sgr A* is 0.15 mJy (dereddened) above the mean level
approximately 34\% of the time. For a somewhat more stringent
higher significant level of 0.3 mJy above the mean, the percentage drops 
to about 23\%.

Dereddened fluxes quoted above were computed using the appropriate
extinction law for the Galactic center (Moneti et al.  2001) and the
Genzel et al.  (2003) extinction value of A(H)=4.3 mag.  These
translate to extinction values for the NICMOS filter bands of
A(F160W)=4.5 mag, A(F187N)=3.5 mag, and A(F190N)=3.4 mag, which then
correspond to corrections factors of 61.9, 24.7, and 23.1.  Applying
these corrections leaves the 1.87 and 1.90$\mu$m fluxes for Sgr~A* at
levels of $\sim$27\%\ and $\sim$7\%, respectively, above the fluxes in
the 1.60 $\mu$m band.  This may suggest that the color of Sgr A* is
red.  However, applying the same corrections to nearby stars, such as
S2 and S0-3, the results are essentially the same as that of Sgr-A*,
namely, the 1.87$\mu$m fluxes are still high relative to the fluxes at
1.60 and 1.90$\mu$m.  This discrepancy in the reddening correction is
likely to be due to a combination of factors.  One is the shape of the
combined spectrum of Sgr~A* and the shoulder of S2, as the wings of S2
cover the position of Sgr~A* .  The other is the diffuse background
emission from stars and ionized gas in the general vicinity of Sgr~A*,
as well as the derivation of the extinction law based on ground-based
filters, which could be different than the NICMOS filter bands.  Due
to these complicating factors, we chose to use the empirically-derived
normalization method described above when comparing fluxes across the
three NICMOS bands.

We have used two different methods to determine the flux of Sgr A* when it 
is flaring. One is to measure directly the peak emission at 1.6$\mu$m 
during the flare to $\approx$0.18 mJy. Using a reddening correction of 
about a factor of 62, this would translate to $\sim$10.9 mJy.  Since we 
have used an aperture radius of only 2 pixels, we are missing a very 
significant fraction of the total signal coming from Sgr A*. In addition, 
the contamination by S2 will clearly add to the measured flux of Sgr A*. 
Not only are we not measuring all the flux from Sgr A* using our 2-pixel 
radius aperture, but more importantly, we're getting a large (but unknown) 
amount of contamination from other sources like S2. 
The second method is 
to determine the relative increase in measured flux which can be safely 
attributed to Sgr A* (since we assume that the other contaminating sources 
like S2 don't vary). The increase in 1.6$\mu$m emission that we have 
observed from Sgr~A* during flare events is $\sim$0.03 mJy, which 
corresponds to a dereddened flux of $\sim$1.8 mJy. Based on photometry of 
stars in the field, we have derived an aperture correction factor of 
$\sim$2.3, which will correct the fluxes measured in our 2-pixel radius 
aperture up to the total flux for a point source. Thus, the increase in 
Sgr A* flux during a flare increases  to a dereddened value of 
$\sim$4.3 mJy. 
Assuming that all of the increase comes from just Sgr-A*, and then adding 
that increase to the 2.8 mJy quiescent flux (Genzel et al. 2003), then we 
measure a peak dereddened H-band flux of $\sim$7.5 mJy during a flare. 
However, recent detection of 1.3 mJy dereddened flux at 3.8$\mu$m from 
Sgr~A* (Ghez et al. 2005) is lower than the lowest flux at H band that had 
been reported earlier (Ghez et al. 2005). This implies that the flux of 
Sgr A* may be fluctuating constantly and there is no quiescent state in 
near-IR band. Given the level of uncertainties involved in both 
techniques, we have used the first method of measuring the peak flux 
which is  adopted as the true flux of Sgr A* for the rest of the 
paper. If the second method  is 
used, the peak flux of Sgr A* should be lowered by a factor of $\sim$0.7.

We  note that the total amount of time 
that flare activity has been detected is roughly 30--40\%\ of the 
total observation time.
It is remarkable that Sgr~A* is active at these levels for 
such a high 
fraction of the time at near-IR wavelengths, especially when compared 
to its X-ray activity, which has been detected on 
the average of once a day or about 1.4 to 5\% of the observing time 
depending on 
different instruments (Baganoff et al. 
2003; Belanger et al. 2005a). 
In fact, over the course of one week of observations in 2004, XMM 
detected only two clusters of X-ray flares. 
Recent detection of 1.3 mJy dereddened  flux at 3.8$\mu$m from 
Sgr~A* is lower than the  lowest flux at H band that had been reported 
earlier (Ghez et al. 2005). This measurement when combined with 
our variability analysis 
is  consistent with the conclusion that 
the near-IR flux of Sgr A*  due to flare activity is 
fluctuating constantly  at a  low  level and that 
there is no quiescent flux.

Figure 3b shows a  histogram plot of the detected flares and the noise 
as well as the simultaneous 2-Gaussian fit to both the
noise and the flares. In the plot the dotted  lines are the individual
Gaussians, while the thick dashed line is the sum of the two. 
The variations near zero is best fitted with a 
Gaussian, which is expected  from random
noise in the observations, while the positive half of the
histogram shows a tail extending out to $\sim$2 mJy above the mean,
which represents the various flare detections.
The
flux values  are dereddened values
within the  4-pixel diameter photometric aperture at 1.60$\mu$m.

The "flux variation" values were computed by first computing the mean 
F160W flux within one of our "quiescent" time windows and then subtracting 
this quiescent value from all the F160W values in all time periods. So 
these values represent the increase in flux relative to the mean 
quiescent. The parameters of the fitted Gaussian for the flares is 10.9, 
0.47$\pm$0.3 mJy, 1.04$\pm$0.5 mJy corresponding to the amplitude, center 
and FWHM, respectively.  The total area of the individual Gaussians are 
26.1 and 12.0 which gives the percentage of the area of the flare 
Gaussian, relative to the total of the two, to be $\sim$31\%.  This is 
consistent with our previous estimate that flares occupy 30-40\% of the 
observing time.

A mean quiescent 1.6$\mu$m flux of 0.15 mJy (observed)
corresponds to a dereddened flux of $\sim$9.3 mJy within a 4-pixel 
diameter aperture. The total flux for a typical flare event (which gives
an increase of 0.47 mJy)  would be $\sim$9.8 mJy. But of course all of
these measurements refer to the amount of flux collected in a 
4-pixel diameter aperture, which includes some contribution from
S2 star and at the same time does not include all the flux of Sgr A*.
If we include  the increase  associated with a typical flare,
which excludes any contribution from S2, and apply the aperture
correction factor of 2.4 which accounts  for 
the amount of missing light from Sgr A*, then the typical flux of  
of 0.47 mJy corresponds 
to a value of 1.13 mJy. If we then use the quiescent flux of Sgr A* 
at H-band (Genzel et al. 2003), the absolute flux 
a  typical flare at 1.6$\mu$m is estimated to be  $\sim$3.9 mJy. 
The energy output per event from a typical flare  with a duration of 
30 minutes is then estimated to be $\sim$10$^{38}$ ergs.
The Gaussian nature of the flare histogram suggests that this estimate 
corresponds to the  characteristic 
energy scale of the accelerating events (If we use a typical flux of a 
flare $\sim$9.8 mJy,  then the  energy scale increases by a factor of 
2.5).

In terms of power-law versus Gaussian fit, the power-law fit to the
flare portion only gave a $\chi^2$=2.6 and rms=1.6, while the
Gaussian fit to the flare part only gives a $\chi^2$=1.6 and
rms=1.2 (better in both). With the limited data we have, 
we believe that it is   difficult to fit a power-law to the flare portion along
with a Gaussian to the noise peak at zero flux, because the power-law
fit continues to rise dramatically as it approaches to zero flux, which
then swamps the noise portion centered at zero. 
 
During the relatively quiescent periods of our observations,  the
observed 1.6 $\mu$m fluxes have a 1$\sigma$ level of $\sim$0.002-0.003 mJy. 
Looking at the periods during which
we detected obvious flares, 
an increase of $\sim$0.005 mJy is noted. 
This is about  2$\sigma$  relative to the observation-to-observation
scatter quoted above  ($\sim$0.002 mJy).
To compare these values to the ground-based data 
using the same
reddening correction as  Genzel et al. (2003),
our 1-$\sigma$  scatter would be about $\sim$0.15 mJy at
1.6 $\mu$m, with our  weakest detected  flares having 
a flux $\sim$0.3 mJy at 1.6 $\mu$m. Genzel et al. report
H-band  weakest detectable variability at  about the 0.6 mJy level.
Thus, the HST  1-$\sigma$  level is about a factor of 4 better 
and the weakest detectable flares about a factor of 2 better
than ground-based observations. 

\subsubsection{Power Spectrum  Analysis}

Motivated by the report of a 17-minute periodic signal from Sgr 
A* in near-IR wavelengths (Genzel et al. 2003), the power 
spectra of our unevenly-spaced near-IR flares were measured using 
the Lomb-Scargle periodogram (e.g., Scargle 1982). There are 
certain 
possible artificial signals
 that should be considered in periodicity analysis of 
HST data.
One is the  22-minute  
cycle of the three filters of  NICMOS observations. 
 In addition, the orbital period of HST 
is 92 minutes, 46 minutes of which 
no observation can be made 
due to the Earth's occultation. Thus, any signals at the 
frequencies corresponding to the inverse of these periods,  
or their harmonics, are of doubtful significance. In spite of these 
limitations the data is sampled and characterized well for the 
periodic analysis.  In order to determine the significance of 
power at a given frequency, we employed a Monte Carlo technique 
to simulate the power-law noise following an algorithm that has 
been applied to different data sets (Timmer \& K\"onig 1995; 
Mauerhan et al. 2005). 5000 artificial light 
curves were constructed for each time segment.  Each simulated 
light curve contained  red noise, following P($f) \propto f^{-\beta}$, and was 
forced to have the same variance and sampling as the original  data.  

Figures 4a,b show the light curves, power spectra, and envelopes of simulated power 
spectra for the flares during the 2nd and 4th observing time windows. The flare 
activity with very weak signal-to-noise ratio at the end of the 4th observing window 
was not included in the power spectrum analysis.  The flares shown in Figures 4a,b 
are separated by about two days from each other and the temporal and spatial 
behavior of of their light curves are similar. Dashed curves on each figure 
indicate the envelope below which 99\% (higher curve), 95\% (middle curve), and 50\% (lower 
curve) 
of the simulated power spectra lie.
These curves show ripples which incorporate information about the
sampling properties of the lightcurves.

The vertical lines represent the period of an HST orbit and 
the period at which the three observing filters were cycled. The 
only  signals which appear  to be slightly above the 
99\% light curve of the simulated power spectrum are
 at  0.55$\pm$0.03 hours, or 33$\pm$2 minutes. 
The power spectrum of the sixth observing window  
shows similar significance near 33 minutes, but also shows similar
significance at other periods near the minima in the
simulated lightcurves.  We interpret this to suggest that the power in
the sixth observation is not well-modeled as red noise.

We compared the power 
spectrum of the averaged data 
from three observing windows using a range of $\beta$ from 1 to 3. 
The choice of  $\beta$=2 
shows the best overall match between 
the line enclosing 50\% of the simulated power spectra and the actual
power spectrum.  A  
$\beta$ of 3 is not too different in the overall fit to that of $\beta=2$. For the 
choice of $\beta$=1, significant power at longer time scales becomes    
 apparent. However, the significance of longer periods 
in the power spectrum disappears when $\beta=2$ was selected, thus 
we take $\beta$=2 to be the optimal value for our analysis. 

The only 
signal that reaches a 99\% significance level 
is the 33-minute time scale. 
This time scale  is about twice the 
17-minute time scale that earlier ground-based observations 
reported (Genzel et al. 2003).  
There is no evidence for any 
periodicity  at 17
minutes in our data. 
The time scale  of about 
33 minutes roughly agrees with the 
timescales on 
which the flares rise and decay. 
Similarly, the power spectrum analysis of X-ray data show 
several periodicities, one of  which falls within the 33-minute time 
scale of HST data (Aschenbach et al. 2004; Aschenbach 2005). 
However, we are doubtful whether 
 this signal indicates a real periodicity.
This signal is only slightly above the 
noise spectrum in all of our simulations and is at best a marginal
result. 
It is clear that any possible periodicities 
need to be confirmed with future HST
observations with better time coverage and more regular time spacing. 
Given that the low-level amplitude variability  that is detected here 
with {\it HST} data is significantly better than what can be 
detected with ground-based telescopes,  
additional HST observations are still required 
to fully understand the 
power spectrum behavior of near-IR flares from Sgr A*.

\subsection {Submillimeter Wavelengths: CSO and SMT}

\subsubsection {CSO Observations at 350, 450, 850 $\mu$m}

Using CSO with SHARC II, Sgr A* was monitored at 450 and 850 $\mu$m in 
both observing epochs (Dowell et al. 2004). 
Within the 2 arcminute field of view of the CSO images,
a central point source 
coincident with Sgr A* is visible at 450 and 850 $\mu$m wavelengths 
having spatial resolutions of 11$''$ and 21$''$, respectively.  Figure~5a 
shows the light curves of 
Sgr A* in the second observing 
epoch with 1$\sigma$ error bars corresponding to 20min of integration. 
The 1$\sigma$  error bars are noise
and relative calibration uncertainty added in quadrature.  Absolute
calibration accuracy is about 30\% (95\% confidence).
During the first epoch, when a transient source appeared a 
few arcseconds away from Sgr A*,
no significant 
variability was  detected. The flux 
density of Sgr A* at 850 $\mu$m is consistent with the SMT flux 
measurement of Sgr A* on March 28, 2004, as discussed below.  During 
this epoch, Sgr A* was 
also observed briefly at 350 $\mu$m on April 1 and showed a flux density 
of 2.7$\pm$0.8 Jy.

The light curve of Sgr A* in the second epoch, presented in 
Figure 5a, shows only $\sim$25\% variability at 450 $\mu$m. 
However, the flux density appears to vary at 850$\mu$m in the 
range between 2.7 and 4.6 Jy over the course of this observing 
campaign. 
 Since the CSO slews slowly, and we need all of the Sgr A*
signal-to-noise, we only observe calibrators hourly.  The
hourly flux of the calibrators 
as a function of  atmospheric opacity 
shows  $\sim$30\% peak-to-peak uncertainty for a
particular calibration source and 
a 10\% relative calibration uncertainty (1$\sigma$) for the CSO 850 micron
data.

We note the presence of  remarkable flare 
activity at 850 $\mu$m on the last day of the observation during 
which a peak flux density of 4.6 Jy was detected with a S/N 
= 5.4. The reality of this flare activity is best demonstrated 
in a map, shown in Figure 5b, which  shows  the 
850$\mu$m flux from well-known 
diffuse features associated with the  southern 
arm of the circumnuclear ring  remaining constant, while the 
emission from Sgr A* rises  to 4.6 Jy during the active 
period.
The feature of next highest significance 
after Sgr A* in the subtracted  map
showing the variable sources  
is consistent with noise with   S/N = 2.5.

\subsubsection {SMT  Observations at 870 $\mu$m}

Sgr A$^*$ was monitored in the 870$\mu$m atmospheric window
using the MPIfR 19 channel bolometer on the Arizona Radio
Observatory (ARO) 10m HHT telescope (Baars et al. 1999).
The array covers a total area of 200$''$ on the sky, with the 19 
channels (of 23$''$ HPBW) arranged in two concentric hexagons around 
the central channel, with an average separation of 50$''$ between any 
adjacent channels. The bolometer
is optimized for operations in the 310-380 GHz (970-790 $\mu$m) region,
with a maximum sensitivity peaking at 340 GHz near 870 $\mu$m. 

The observations were carried out in the first epoch during the period 
March 28-30th, 2004  between 11-16h UT.
Variations of the atmospheric optical depth at 870$\mu$m were measured
by straddling all observations with skydips. The absolute gain of the
bolometer channels was measured by observing the planet Uranus at the
end of each run. A secondary flux calibrator, i.e. NRAO 530, was 
observed 
to check the stability and repeatability of the measurements. 
All observations were carried out with a
chopping sub-reflector at 4Hz and with total beam-throws in the range 
120$''-180''$, depending on a number of factors such as weather 
conditions 
and elevation. 


As already noted above, dust around Sgr A$^*$ is clearly 
contaminating our measurements at a resolution of 23$''$. Due to the 
complexity of this field, the only possibility to try to recover the 
uncontaminated flux is to fit several components to the brightness 
distribution, assuming that in the central position, there is an 
unresolved source, surrounded by an extended smoother distribution. 
We measured the average brightness in concentric rings (of 8$''$ width) 
centered on Sgr A$^*$ in the radial distance range 0-80$''$. The 
averaged radial profile was then fitted with several composite 
functions, but always included  a point source with a PSF of the 
order of the beam-size. The best fit for both the central component 
and a broader and smoother outer structure gives a central (i.e., 
Sgr A$^*$) flux of 4.0$\pm$0.2Jy in the first day of observation 
on March 28, 2004. 
The CSO source flux fitting, as described earlier, 
and HHT fitting  are essentially the same.

Due to bad weather, the scatter in the measured flux of the 
calibrator NRAO 530 and Sgr A* was  high in the second and third days 
of the run. Thus, the measurements reported here 
are  achieved only for the first day 
with the photometric precision 
 $\leq$12$\%$ for  the calibrator.  
The flux of NRAO 530 at 870$\mu$m during this  observation was 
1.2$\pm$0.1 Jy.

\subsection {Radio Wavelengths: NMA,  BIMA, VLA \& ATCA}

\subsubsection { NMA Observations at 2 \&3mm}

NMA was used  in the first observing epoch to observe
Sgr A* at 3 mm (90 GHz) and 2 mm (134 GHz), as part of a long-term
monitoring
campaign  (Tsutsumi, Miyazaki \& Tsuboi 2005).
The 2 and 3 mm flux density were measured to be 1.8$\pm$0.4 and
2.0$\pm$0.3 Jy on March 31 and April
1, 2004, respectively,  during 2:30-22:15 UT.  These authors had
also reported  a  flux density of 2.6$\pm$0.5 Jy
at 2 mm on
March 6, 2004.  This observation took place when
a  radio and X-ray  transient near Sgr A* was active.
Thus, it is quite possible that the  2 mm emission toward
Sgr A* is not  part of   a flare activity from Sgr A* but
rather due to decaying emission from a radio/X-ray transient
which was first detected by  XMM and VLA on March 28, 2004.
                                                                                                                    
\subsubsection { BIMA  Observations at 3mm}

Using nine telescopes, BIMA observed Sgr A* at 3 mm (85 GHz, average
of two sidebands at 82.9 and 86.3 GHz) for five days
between March 28 and April 1, 2004 during 11:10-15:30 UT .  Detailed
time variability analysis is given elsewhere (Roberts et
al. 2005). The flux densities on March 28 and April 1 show average
values of 1.82$\pm$0.16 and 1.87$\pm$0.14 at $\sim$3 mm, respectively.
These values are consistent with the NMA flux values within errors.
No significant hourly variability was detected.
                                                                                                                    
The presence of the transient X-ray/radio source a few
arcseconds south of Sgr A* during this epoch complicates time
variability analysis of BIMA data since the relatively large 
synthesized beam
(8\dasec2 $\times$ 2\dasec6) 
changes during the course of the
observation.  Thus, as the beam rotates throughout an observation,
flux included from Sgr A West and the radio transient may contaminate
the measured flux of Sgr A*.

\subsubsection {VLA  Observations at 7mm}

Using the VLA, Sgr A* was observed at 7mm (43 GHz) in the first and
second observing epochs.  In each epoch, observations were carried out
on four consecutive days, with an average temporal coverage of
about 4 hr per day.  In order to calibrate out rapid atmospheric
changes, these observations used a new fast switching technique for
the first time to observe time variability of Sgr A*.  
Briefly, these observations used the same calibrators
(3C286, NRAO 530 and 1820-254).  The fast switching mode rapidly
alternated between Sgr A* (90sec) and the calibrator 1820-254 (30sec).
Tipping scans were included every 30 min to measure and correct for
the atmosphere opacity. In addition, pointing was done by observing
NRAO 530.  After applying high frequency calibration, the flux of Sgr
A* was determined by fitting a point source in the {\it uv} plane
 ($>$100 k$\lambda$).  As a check,  the variability data were 
also analyzed
in the image plane, which gave similar results.
                                                                                                                    
The results of the analysis at 7mm clearly indicate a 5-10\%
variability on hourly time scales,  in almost all the observing
runs. A power spectrum  analysis, similar to the statistical analysis 
of near-IR data presented above, was also done at 7mm. 
 Figure 6a
shows typical light curves of NRAO 530 and Sgr A* in the top two
panels at 7mm. 
Similar behavior is found in a 
number of
observations during 7mm observations in both epochs.  
It is clear that
the light curve starts with a peak (or that the peak preceded the
beginning of the observation) followed by a decay with a  duration of 30 
minutes 
to a quiescent level lasting for about 2.5 hours.  

                                                                                                                    
\subsubsection {ATCA Observations at 1.7 and 1.5cm}

At the  ATCA, we used a similar  observing technique to that of our VLA
observations, involving fast 
switching between the calibrator and Sgr A*  simultaneously at
1.7 (17.6 GHz) and 1.5 cm (19.5 GHz). Unlike ground based northern
hemisphere observatories that can observe Sgr A* for only 5 
hours a
day (such as the VLA), ATCA observed Sgr A* for 4 $\times$ 12 hours
in the first epoch.  
 In spite of the
possible contamination of variable flux due to interstellar
scintillation toward Sgr A* at longer wavelengths, 
similar variations in both 7 mm and 1.5 cm are detected. 
Figure 6b shows the light curve of Sgr A* and the corresponding 
calibrator during a 12-hour observation with ATCA at 1.7cm. 
The increase in the flux of Sgr A* is seen with a rise and fall time 
scale of about 2 hours.
The 1.5, 1.7 cm
and 7 mm variability analysis is not inconsistent with the 
time scale at
which significant power has been reported  at 3 mm (Mauerhan et
al. 2005).  Furthermore, the  time scales for rise and fall  
of flares in radio wavelengths are longer than in the near-IR wavelengths
discussed above.
                                                                                                                    
\section {Correlation Study}
\subsection{Epoch 1}

Figure 7 shows the simultaneous light curves of Sgr A* during the
first epoch in March 2004 based on observations made with XMM, CSO at
450 and 850 $\mu$m, BIMA at 3 mm and VLA at 7 mm.  The flux of Sgr A*
remained constant in submillimeter and 
millimeter wavelengths throughout the first epoch, while we
observed an X-ray flare (top panel) at the end of the XMM observations and
hourly variations in radio wavelengths (bottom panel) at a level
10-20\%.  
This implies that the contamination from the radio and X-ray transient
CXOGCJ174540.0-290031, which is located a few arcseconds from Sgr A*, is
minimal, thus the measured fluxes should represent the quiescent 
flux of Sgr A*. These data are used to make a spectrum of Sgr 
A*, as discussed in section 5. As for the X-ray flare, there were 
no simultaneous observations with other instruments during  the period 
in which the 
X-ray flare took place. Thus, we can not 
state if there were any variability 
at other wavelengths during the X-ray flare in this epoch.

\subsection {Epoch 2}

Figure 8 shows the simultaneous light curve of Sgr A* based on the
second epoch of observations using XMM, HST, CSO and VLA.
Porquet et al. (2005) noted clear 8-hour periodic dips 
due to the eclipses of the  transient as seen clearly in the XMM 
light curve. 
Sgr A*
shows clear variability at near-IR and submillimeter wavelengths, as
discussed below.

One of the most exciting results of this observing campaign is the
detection of a cluster of near-IR flares in the second observing
window which appears to have an X-ray counterpart.  The long temporal
coverage of XMM-Newton and HST observations have led to the detection
of a simultaneous flare in both bands.  However, the rest of the near-IR
flares detected in the fourth and sixth observing windows (see Figure
3) show no X-ray counterparts at the level that could be detected with
XMM. The two brightest near-IR flares in the second and fourth observing
windows are separated by roughly two days and appear to show similar
temporal and spatial behaviors.  Figure 9 shows the simultaneous
near-IR and X-ray emission with an amplitude increase of $\sim$15\% and 
 100\% for the peak emission, respectively. 
We believe that these flares are
associated with each other for the following reasons.  First, X-ray
and near-IR flares are known to occur 
from Sgr~A* as previous
high resolution X-ray and near-IR observations have pinpointed the
origin of the flare emission. Although near-IR flares 
could be active up to 40\% of time, the X-ray flares are generally 
rare with a 1\%  probablity of occurance  based on a week of 
observation with XMM. Second, although the  chance coincidence for a near-IR flare to 
have an X-ray counterpart could be high but what is 
clear from Figure 9 is the way that near-IR and X-ray flares track
each other on a short time. 
Both  the near-IR and X-ray flares
show similar morphology in their light curves as well as similar
duration with no apparent delay. This  leads us to believe that both 
flares
come from the same region close to the event horizon of Sgr A*. 
The X-ray light curve shows a double peaked maximum  flare near Day 
155.95 which appears to be 
remarkably in phase with the near-IR strongest double peaked flares, 
though with different amplitudes. We can also note similar trend 
in the sub-flares noted near Day 155.9 in Figure 9 where they show similar 
phase but different amplitudes. 
 Lastly, since
X-ray flares occur on the average once a day, the lack of X-ray
counterparts to other near-IR flares indicates clearly that not
all near-IR flares have X-ray counterparts.  
This fact has important
implications on the emission mechanism, as described below.

With the exception of the September 4, 2004 observation toward  the 
end 
of the second
observing campaign, the large error bars of the submillimeter data do not
allow us to determine short time scale variability in this wavelength domain
with high confidence.  We notice a significant increase in the
850$\mu$m emission about 22 hours after the simultaneous X-ray/near-IR
flare took place, as seen in Figure 8.  We also note the highest
850$\mu$m flux in this campaign 4.62$\pm$0.33 Jy which is 
detected 
toward the
end of the submillimeter observations.  
This corresponds to a  5.4$\sigma$ increase of 850$\mu$m flux.

Figure 10
shows simultaneous light curves of Sgr A* at 850$\mu$m and near-IR
wavelengths.  
The strongest near-IR flare occurred at the beginning of the 6th
observing window with a decay time of about 40 minutes followed by the
second flare about 200 minutes later with a decay time of about 20
minutes.  The submillimeter light curve shows a peak about 
160
minutes after the strongest near-IR flare that was detected in the
second campaign. 
The duration of the submillimeter flare is about two hours. 
Given that there is no near-IR data during one half of 
every HST orbit and that 
the 850$\mu$m data were  sampled every 20 minutes compared to
32sec sampling rate in near-IR wavelengths, it is not clear 
whether  the submillimeter data is correlated simultaneously 
with the 
second bright 
near-IR  flare, or is produced by the first near-IR flare with a delay
of 160 minutes, as seen in Figure 10.   
What is significant is that  submillimeter data  suggests
that
the 850$\mu$m emission is variable and is correlated
with the near-IR data. Using optical depth and polarization arguments, we 
argue below that the submillimeter 
and near-IR flares are simultaneous.

\section {Emission Mechanism}
\subsection {X-ray and Near-IR Emission}

Theoretical studies of accretion flow near Sgr A* show that 
the flare emission in near-IR and X-rays  
 can be accounted for in terms of the acceleration of particles to high energies, producing 
synchrotron emission as well as ICS  (e.g., Markoff et al. 2001; Liu \& Melia 2001; Yuan, Markoff \& 
Falcke 2002; 
 Yuan, Quataert \& Narayan 2003, 2004). 
Observationally, the near-IR flares are known to be due to synchrotron emission based on spectral index 
and polarization measurements 
(e.g., Genzel
et al.  2003 and references therein). 
We argue that the X-ray counterparts to the near-IR flares are unlikely to be produced
by synchrotron radiation in the typical $\sim10$\,G magnetic field
inferred for the disk in Sgr A*  for two reasons.  First,
emission at 10\,keV would be produced by 100\,GeV electrons, which
have a synchrotron loss time of only 20\,seconds, whereas individual
X-ray flares rise and decay on much longer time scales.  Second, the
observed spectral index of the X-ray counterpart, $\alpha=0.6$ ($S_\nu
\propto \nu^{-\alpha=-0.6}$), does not match the near-IR to X-ray 
spectral index.  The
observed X-ray 2-10 keV flux 6$\times10^{-12}$ erg cm$^{-2}$ s$^{-1}$
corresponds to a differential flux of 2$\times10^{-12}$ erg cm$^2$
s$^{-1}$ keV$^{-1}$ (0.83 $\mu$Jy) at 1 keV. The extinction-corrected
(for $A_H=4.5$\,mag) peak flux density of the near-IR (1.6$\mu$m)
flare is $\sim$10.9 mJy.  The spectral index between X-ray and near-IR
is 1.3, far steeper than the index of 0.6 determined for the X-ray
spectrum.

Instead, we favor an inverse Compton model 
for the X-ray emission,
which naturally produces a strong correlation with the near-IR flares.
In this picture, submillimeter photons are upscattered to X-ray
energies by the electrons responsible for the near-IR synchrotron
radiation.  The fractional variability at submillimeter wavelengths
is less than 20\%, so we first consider quiescent submillimeter
photons scattering off the variable population of GeV electrons that
emit in the near-IR wavelengths.
In the ICS picture, the spectral index of the near-IR flare must 
match that of the X-ray
counterpart, i.e. $\alpha$ = 0.6.  Unfortunately, we were not able to
determine the spectral index of near-IR flares.  
Recent measurements of the spectral index of near-IR 
flares appear to vary considerably ranging between 0.5 
to 4 (Eisenhauer et al. 2005; Ghez et al. 2005).  The 
de-reddened peak flux of 10.9 mJy (or 7.5 mJy from  the
relative flux measurement described in section 2.2.2) with 
a spectral index of 0.6 is  consistent  with a picture that 
 blighter near-IR flares  have harder spectral index
(Eisenhauer et al. 2005; Ghez et al. 2005).

Assuming an 
electron
spectrum extending from 3\,GeV down to 10\,MeV and neglecting the
energy density of protons, the equipartition magnetic field is 11\,G,
with equipartition electron and magnetic field energy densities of
$\sim$5 erg cm$^{-3}$.  The electrons emitting synchrotron at
1.6$\mu$m then have typical energies of 1.0 GeV and a loss time of
35\,min.
1\,GeV electrons will Compton scatter 850\,$\mu$m photons up to
7.8\,keV, so as the peak of the emission spectrum of Sgr A* falls in
the submillimeter regime, as it is natural to consider the upscattering
of the quiescent submillimeter radiation field close to Sgr
A*.  We assume that this submillimeter emission arises
from a source diameter of 10 
Schwarzschild radii (R$_{sch}$), or
0.7\,AU (adopting a black hole mass of 3.7$\times10^{6}$ $\msol$).
In order to get the X-ray flux, we need the spectrum of the seed photons
which is not known. We make an assumption that   
the measured submillimeter flux  
 (4 Jy at 850 $\mu$m), and the product of the spectrum of the near-IR emitting 
particles and submillimeter flux $\nu^{0.6} F_{\nu}$, are 
 of the same order over a decade in frequency.
The predicted ICS X-ray flux for this simple model is
$1.2\times10^{-12}$ erg cm$^{-2}$ s$^{-1}$ keV$^{-1}$, roughly half of
the observed flux.  

The second case we  consider to explain the origin of X-ray 
emission is that  
near-IR photons  scatter off the 
population of $\sim$50  MeV electrons that
emit in submillimeter  wavelengths.
If synchrotron emission from a population of
lower-energy ($\sim 50$\,MeV) electrons in a similar source region
(diameter $\sim 10$\,R$_{sch}$, $B\sim 10$\,G) is responsible for the
quiescent emission at submillimeter wavelengths, then upscattering of
the flare's near-IR emission by this population will produce a similar
contribution to the flux of the X-ray counterpart, and the predicted
net X-ray flux  $\sim2.4\times10^{-12}$ erg cm$^2$ s$^{-1}$ keV$^{-1}$
is similar to that observed. 

The two physical pictures of ICS described above produce similar X-ray 
flux within  the inner 
diameter $\sim 10$\,R$_{sch}$, $B\sim 10$\,G, and therefore 
cannot be  distinguished from each other. 
On the other hand,  if the near-IR flares  arise from a   region 
smaller than 
that of the 
quiescent submillimeter seed photons, then 
the first case, in which the  quiescent submillimeter
photons scatter off GeV electrons that
emit in the near-IR, is a more likely mechanism to 
produce X-ray flares. 

The lack of an X-ray counterpart to every detected near-IR flare
can be explained 
naturally in the ICS picture presented here.  It can be
understood in terms of variability in the magnetic field strength or
spectral index of the relativistic particles, two important parameters
that determine the 
relationship between the near-IR and ICS X-ray flux.  
A large variation of the spectral index in near-IR 
wavelengths has been 
observed (Ghez et al. 2005; Eisenhauer et al. 2005).  
Figure 11a shows the ratio of the fluxes at 1 keV and 1.6 $\mu$m
against the spectral index for different values of the magnetic field.
Note that there is a minimum field set by requiring the field energy
density to be similar to or larger than the relativistic particle
energy.  
If, as is likely, the magnetic field is ultimately responsible 
for the acceleration of the
relativistic particles, then the field pressure must be stronger 
or equal to the particle
energy density so that the particles are confined by the field during 
the acceleration process.
It is clear that hardening (flattening) of the spectral index
and/or increasing the magnetic field reduces the X-ray flux at 1 keV
relative to the near-IR flux.  
 On the other hand
softening (steepening) the spectrum can produce strong X-ray flares.
This occurs 
because a higher fraction of
relativistic particles have lower energies and are, therefore,
available to upscatter the submillimeter photons. 
This is consistent with the fact that the strongest X-ray flare that
has been detected from Sgr A* shows the softest (steepest) spectral
index (Porquet et al.  2003). Moreover, the sub-flares presented in 
near-IR and in X-rays, as shown in Figure 9, 
appear to indicate that the ratio of 
X-ray to near-IR flux (S$_X$ to S$_H$)  
varies in two sets of double-peaked flares, as  described 
earlier. We  note an 
X-ray spike at   155.905 days has  a 1.90 $\mu$m  (red color) 
counterpart. 
The
preceding 1.87 $\mu$m (green color)  data points are all steadily 
decreasing from the
previous flare, but then the 1.90$\mu$m suddenly increases 
 up to at least at a level of $\sim3\sigma$. 

The flux  ratio corresponding to  the peak X-ray flare (Figure 9) is high as 
it argues that the flare has either a soft spectral index and/or a low magnetic field. 
Since the strongest X-ray flare that has been detected thus far  has the steepest spectrum 
(Porquet et al. 
2003), we believe that the observed variation of the  flux ratio 
in Sgr A* is due to   the variation of the spectral index
of individual near-IR flares. Since most of the observed X-ray sub-flares are 
clustered temporally, it is plausible to consider that they all arise from the same
location  
in the disk. This implies that the the strength of the magnetic field does 
not vary between sub-flares.

\subsection{Submillimeter and Near-IR Emission}

As discussed earlier, we cannot determine whether the submillimeter
flare at 850$\mu$m is correlated with a time delay of 160 minutes or
is simultaneous with the detected near-IR flares (see Fig.  10).
Considering that near-IR flares are relatively  continuous 
with up to   40\% probability and that the near-IR and submillimter flares are  due
to chance coincidence, 
the evidence for a delayed or simultaneous correlation between  these
two flares  is not clear.  However, spectral index measurements 
in submillimeter domain as well as a jump in the polarization 
position angle in submillimeter wavelengths suggest that the transition from 
optically  thick to 
thin regime occurs near 850 and 450 $\mu$m  wavelengths (e.g.,
 Aitken et al. et al. 2000; Agol 
2000; Melia et al. 2000;  D. Marrone, private communication).
If so, it is reasonable to consider  
that the near-IR and submillimeter flares are simultaneous with no time delay 
and these flares are generated by synchrotron emission from the same population of
electrons.  
Comparing the 
peak flux 
densities of 11 mJy and 0.6 Jy at
1.6$\mu$m and 850$\mu$m, respectively, gives a spectral index
$\alpha\sim 0.64$ (If we use a relative flux of 7.6 mJy at 1.6$\mu$m,
the $\alpha\sim 0.7$). 
This assumes that 
the population of synchrotron
emitting particles in near-IR wavelengths with typical energies of
$\sim$1 GeV could extend down to energies of $\sim50$ MeV. A
low-energy cutoff of 10 MeV was assumed in the previous section to
estimate the X-ray flux due to ICS of seed photons.  In this picture,
the enhanced submillimeter emission, like near-IR emission, is mainly
due to synchrotron and arises  from the inner 10R$_{sch}$ of Sgr
A$^*$ with a magnetic field of 10G. 
Similar to the argument made in the previous section, 
the lack of
one-to-one correlation between near-IR and submillimeter flares could
be due to the varying energy spectrum of the particles generating
near-IR flares. 
The hard(flat) spectrum of radiating particles will be less effective
in the production of submillimeter emission whereas the soft (steep) spectrum
of particles should generate enhanced synchrotron emission at
submillimeter wavelengths.  
This also implies that the variability of steep spectrum near-IR flares  should be
correlated with submillimeter flares.
The synchrotron lifetime of particles
producing 850$\mu$m  is about 12 hours,  which is much longer than the
35\,min time scale for the GeV particles responsible for the near-IR
emission.  Similar argument can also be made for the near-IR flares since 
we detect the rise or  fall time scale of some of the near-IR flares to be about 
ten minutes 
which is shorter than the synchrotron cooling time scale. Therefore we 
conclude that the duration of the
submillimeter and near-IR flaring must be set by dynamical mechanisms
such as adiabatic expansion rather than frequency-dependent processes 
such as synchrotron cooling. The fact that the rise and fall time scale 
of near-IR and submillimeter flare emission is shorter than their corresponding 
synchrotron cooling time scale is consistent with  adiabatic cooling. If we 
make the 
assumption that the 
33-minute time scale detected in near-IR power spectrum analysis is real, this 
argument can 
also be used to rule out 
the possibility that this  time scale  is  due to 
the near-IR cooling time scale.

\subsection{Soft $\gamma$-ray  and Near-IR Emission}
 
As described earlier, a soft $\gamma$-ray INTEGRAL source
IGRJ17456-2901 possibly coincident with Sgr A* has a luminosity of
4.8$\times10^{35}$ erg s$^{-1}$ between 20-120 keV. The spectrum is
fitted by a power law with spectral index $2\pm1$ (Belanger et al.
2005b).  Here, we make the assumption that this source is associated
with Sgr A* and apply the same ICS picture that we argued above for
production of X-ray flares between 2-10 keV. The difference between
the 2-10 keV flares and IGRJ17456-2901 are that the latter source is
detected between 20-120 keV with a steep spectrum and is persistent
with no time variability apparent on the long time scales probed by
the INTEGRAL observations.  Figure 11b shows the predicted peak
luminosity between 20 and 120 keV as a function of the spectral index
of relativistic particles for a given magnetic field.  In contrast to
the result where the softer spectrum of particles produces higher ICS
X-ray flux at 1 keV, the harder spectrum produces higher ICS soft
$\gamma$-ray emission.  Figure 11b shows that the observed luminosity
of 4.8$\times10^{35}$ erg s$^{-1}$ with $\alpha$ = 2 can be matched
well if the magnetic field ranges between 1 and 3 G, however the
observed luminosity must be scaled by at least a factor of three to
account for the likely 30-40\% duty cycle of the near-IR and the
consequent reduction in the time-averaged soft gamma-ray flux. This is 
also consistent with the possibility that much or all of the 
detected soft $\gamma$-ray emission arises from a collection of 
sources within the inner several arcminutes of the Galactic center. 


\section{Simultaneous Multiwavelength Spectrum}

In order to get a simultaneous spectrum of Sgr A*, we used the data
from both epochs of observations.  As pointed out earlier, the first
epoch data probably represents best the quiescent flux of Sgr A*
across its spectrum whereas the flux of Sgr A* includes flare emission
during the second epoch.  Figure 12 shows  power emitted for a given
frequency regime as derived from simultaneous measurements from the
first epoch (in blue solid line).  We have used the mean flux and the
corresponding statistical errors of each measurement for each day of
observations for the first epoch.  Since there were not any near-IR
measurements and no X-ray flare activity, we have added the quiescent
flux of 2.8 and 1.3 mJy at 1.6 and 3.8 $\mu$m, respectively (Genzel et al.  
2003; Ghez et al. 2005) and 20 nJy between 2
and 8 keV (Baganoff et al.  2001) to construct the spectrum shown in
Figure 12.  For illustrative purposes, the hard $\gamma$-ray flux in the TeV 
range (Aharonian et al. 2004)
is also shown in Figure 12.  
The  F$_{\nu} \nu$ spectrum 
peaks at 350 $\mu$m whereas
F$_{\nu}$ peaks at 850 $\mu$m in submillimeter domain.  The flux at
wavelengths between 2 and 3mm as well as between 450 and 850 $\mu$m
appear to be constant as the emission drops rapidly toward radio and
X-ray wavelengths.  The spectrum at near-IR wavelengths is thought to
be consistent with optically thin synchrotron emission whereas the
emission at radio wavelengths is due to optically thick nonthermal
emission.

The spectrum of a flare is also constructed using the flux values in
the observing window when the X-ray/near-IR flare took place and is
presented in Figure 12 as red dotted line.  It is clear that the powers
 emitted in radio and millimeter wavelengths are generally very
similar to each other in both epochs whereas the power is dramatically
changed in near-IR and X-ray wavelengths.  We also note that the slope of
the power generated between X-rays and near-IR wavelengths does not seem
to change during the quiescent and flare phase. However, the flare substructures shown 
in Figure 9 shows clearly that the spectrum between the near-IR to 
X-ray subflares must be varying.  The soft 
and hard  
$\gamma$-ray fluxes based on INTEGRAL  and HESS (Belanger et al. 
2005b; Aharonian et al. 2004) are 
also included in the plot as black dots. 
It is clear that 
F$_{\nu} \nu$ spectrum at TeV is similar to the observed values at 
low energies. This plot also shows that the high flux at  20 keV is 
an upper limit to the flux of Sgr A* because of the contribution 
from confusing  sources within a 13$'$ resolution of INTEGRAL.

The simultaneous near-IR and submillimeter flare emission 
is a natural consequence of optically thin emission. Thus,  both near-IR 
and  submillimeter flare  emission are  nonthermal and no delay is expected
between the near-IR and submillimeter flares in this picture. 
We  also compare the quiescent flux of Sgr A*
with a flux of 2.8 mJy at 1.6$\mu$m with the minimum flux of about 2.7
Jy at 850$\mu$m detected in our two observing campaigns.  The spectral
index that is derived is similar to that derived when a simultaneous
flare activity took place in these wavelength bands, though there is much 
uncertainty as to what the quiescent flux of Sgr A* is in near-IR wavelengths. 
If we use these measurements at face value,  this 
may imply 
that the 
quiescent flux of Sgr A* in near-IR and submillimeter could in principle 
be coupled to each other. The contribution of nonthermal emission to
the quiescent flux of Sgr A* at 
submillimeter wavelength is an 
observational question that needs 
to be determined  in future study of Sgr A*. 

\section{Discussion}

In the context of accretion and outflow models of Sgr A*, a 
variety of synchrotron and ICS mechanisms probing parameter space 
has    been invoked 
to explain the origin of flares from Sgr A*. A detailed analysis of 
previous models of flaring activity, the acceleration mechanism and 
 their comparison 
with the 
simple modeling given here are beyond the scope of 
this 
work.
Many of these models have considered a  broken power law 
distribution or energy cut-offs for the nonthermal particles, or have 
made an assumption of thermal relativistic particles to explain 
the origin of 
submillimeter emission (e.g., Melia \& 
Falcke 2001;  Yuan, Markoff \& 
Falcke 2002; Liu \& Melia 2002; Yuan Quataert \& Narayan 2003, 2004; 
Liu, Petrosian \& Melia 2004; Atoyan \& Dermer (2004); Eckart et al. 2004, 2005;
Goldston, Quataert 
\& Tgumenshchev 2005; Liu, Melia, Petrosian 2005; Guillesen et al. 2005)
The correlated near-IR 
and X-ray flaring which we have observed is consistent with a model in 
which the 
near-IR synchrotron emission is produced by a transient population of 
$\sim$GeV electrons in a $\sim $10\,G magnetic field of size $\sim 
10R_{sch}$. Although ICS and synchrotron mechanisms  
 have   been used in numerous models to explain the quiescent and flare emission from 
Sgr A*  since the 
first discovery of X-ray flare was reported (e.g.,  Baganoff 2001), the  simple 
model of X-ray, near-IR and submillimeter 
emission discussed here is different in that the 
X-ray flux is produced by a roughly equal mix of 
(a) 
near-IR 
photons that are up-scattered by the 50\,MeV particles responsible for the 
quiescent submillimeter emission from Sgr A*, and/or (b) 
submillimeter photons up-scattered from the GeV electron population 
responsible for the near-IR flares. Thus, the degeneracy in these two 
possible mechanisms  can not be removed in this simple model and obviously a
more detailed analysis is needed.
In addition, we predict that the lack of a correlation between 
near-IR and X-ray flare emission can be explained by the variation 
of spectral index and/or the magnetic fields. The variation of these 
parameters in the context of the stochastic acceleration model 
of flaring events has also  been explored recently  
(Liu, Melia and Petrosian 
2005; Gillesen et al. 2005). 
 
The similar durations of the submillimeter and near-IR flares imply 
that the transient population of relativistic electrons loses energy
by a dynamical mechanism such as adiabatic expansion rather than
frequency-dependent processes such as synchrotron cooling.  
The dynamical time scale 1/$\Omega$ (where $\Omega$ is the rotational 
angular
frequency) is the natural expansion time scale of a build up of pressure.
This is because vertical hydrostatic equilibrium for the disc at a given
radius is the same as the dynamical time scale.  In other words, the time
for a sound wave to run vertically across the disc, h/c$_s$ = 1/$\Omega$.
The 30--40
minute time scale can then be identified with the accretion disk's
orbital period  at the location of the emission region, 
yielding
an estimate of $3.1-3.8\,R_{sch}$ for the disc radius where the flaring
is taking place.  This estimate has assumed 
that the black hole is non-rotating (a/M = 0). 
Thus, the orbiting gas corresponding to this period has 
 a radius of 3.3
R$_{sch}$  which  is greater than the size of the last stable orbit. 
Assuming that the significant power at  33-minute time scale is real,
it confirms  our source size assumption
in the  simple ICS
model for the X-ray emission. 
If this 
general picture is correct, then more detailed hot-spot modeling 
of the emission from the accreting gas may be able to abstract 
the black hole mass and spin from spot images and light curves 
of the observed flux and polarization (Bromley, Melia \& Liu 2001; 
Melia et al. 2001; Broderick and Loeb 2005a,b).

Assuming the 33-minute duration of most of the near-IR flares is real, 
this time scale is  also comparable with the synchrotron loss time of the 
near-IR-emitting ($\sim 1$\,GeV) electrons in a 10\,G field.  
This time scale is also of the same order as the inferred 
dynamical time scale in the emitting region. This is not surprising
considering that  if 
particles are accelerated in a short initial burst and are confined to a 
blob that subsequently expands on a dynamical time scale, the 
characteristic age of the particles is just the expansion time scale. 
The duration of submillimeter flare presented here appears to be 
slightly longer (roughly one hour), than the duration of  near-IR flares 
(about 20--40minutes) (see also Eckart et 
al. 2005). 
This is consistent with the picture that the blob size  in the context 
of an  outflow from Sgr A* is more compact than in that  
at submillimeter wavelength.  
The 
spectrum of energetic particles should then steepen above the energy for 
which the synchrotron loss time is shorter than the age of the particles, 
i.e.,  in excess of a few GeV. This is consistent with a steepening of the 
flare spectrum at wavelengths shorter than a micron. 

The picture described above implies that flare activity drives mass-loss 
from the disk surface. The near-IR emission is optically thin, so we can 
estimate the mass of relativistic particles in a blob (assuming equal 
numbers of protons and electrons) and the time scale between blob 
ejections. If the typical duration of a flare is 30 minutes and the flares 
are occurring 40\% of the time, the time scale between flare is estimated 
to be $\sim$75 minutes.  Assuming equipartition of particles and field 
with an assumed magnetic field of 11G and using the spectral index of 
near-IR flare $\alpha=0.6$ identical to its X-ray counterpart, the density 
of relativistic electrons is then estimated to be n$_e=3.5\times10^2$ 
cm$^{-3}$ 
(The steepening of the spectral index value to 1 increases particles 
density to 4.6$\times10^2$ cm$^{-3}$.)  The volume of the emitting region 
is estimated to be $785 R_{Sch}^3$.  The mass of a blob is then $\sim 
5\times 10^{15}$\,g if we use a typical flux of 3.9 mJy at 1.6$\mu$m.  
The time-averaged mass-loss rate is estimated to be $\sim 2 \times 
10^{-12} \msol yr^{-1}$.  If thermal gas is also present at a temperature 
of T$\sim5 \times10^9$ K with the same energy density as the field and 
relativistic particles, the total mass-loss due to thermal and nonthermal 
particles increases to $\sim1.3\times10^{-8}$ \msol yr$^{-1}$ (this 
estimate would increase by a factor of 2.5 if we use a flux of 9.3 mJy for 
a typical flare).  Using a temperature of 10$^{11}$ K, this estimate is 
reduced by a factor of 20.  
 It is clear from 
these estimates that the mass-loss 
rate is much less than the Bondi accretion rate based on X-ray 
measurements (Baganoff et al. 2003). Similarly, recent rotation measure 
polarization measurements at submillimeter wavelength place a constraint 
on the accretion rate ranging between 10$^{-6}$ and 10$^{-9}$ \msol 
yr$^{-1}$ (Marrone et al. 2005).

\section{Summary}

We have presented the results of an extensive study of the 
correlation of flare emission from Sgr A* in several different bands. 
On the observational side, we have reported 
the detection of several near-IR flares, two of which showed X-ray and submillimeter 
counterparts. The flare emission in submillimeter wavelength and its apparent 
simultaneity with a near-IR flare are both shown for the fist time.  Also,  remarkable 
 substructures in X-ray and near-IR light curves are noted suggesting 
 that both flares are simultaneous with no time delays. 
What is clear from 
the correlation analysis of near-IR data is that  relativistic electrons 
responsible for near-IR emission are being 
accelerated 
for  a high fraction of the time (30-40\%) having  a wide range of power law indices. 
This is supported by the ratio of flare emission in near-IR to X-rays. 
In addition, the near-IR data shows  a marginal detection of  periodicity  on a time 
scale of $\sim$32 
minutes.  Theoretically, we have used a simple ICS model to explain 
the origin of X-ray and soft $\gamma$-ray emission. 
The mechanism to up-scatter the seed submillimter photons by the GeV 
electrons which produce near-IR synchrotron emission has been used to 
explain the origin of simultaneous near-IR and X-ray flares. 
 We  also explained that the 
submillimeter flare emission is due  to synchrotron emission with relativistic particle 
energies
extending down to $\sim$50 MeV.   Lastly, the equal flare time scale in submillimeter 
and 
near-IR wavelengths implies that the burst of emission  expands and cools  on a 
dynamical time 
scale before 
they leave Sgr A*. We suspect that the simple outflow picture presented here shows some of 
the characteristics that may  take place in micro-quasars such as 
GRS 1915+105 (e.g., Mirabel and Rodriguez 1999).


Acknowledgments: We thank J. Mauerhan and M. Morris for 
providing us with an algorithm to generate the power 
spectrum of noise and 
L. Kirby, J. Bird, and M. Halpern for assistance with the CSO
observations.  We also thank A. Miyazaki for providing us the NMA  data 
prior  publication.



\begin{figure}
\plotfiddle{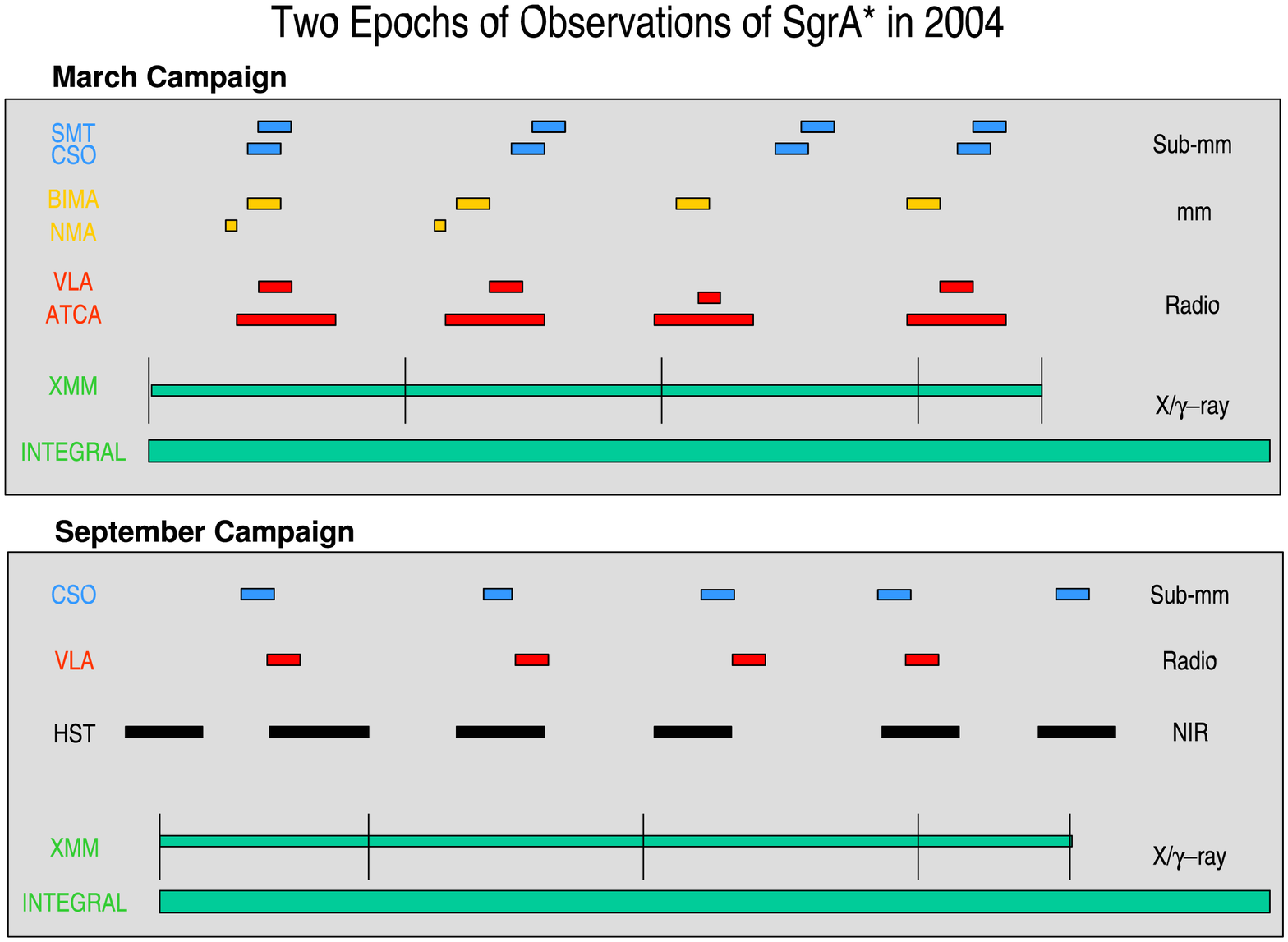}{6in}{0}{70.0}{70.0}{-220}{50}
\caption{ A schemetic diagram showing different telescopes used 
in both observing campaigns. The width of individual observing 
period is not scaled. }\end{figure}

\begin{figure}  
\plotfiddle{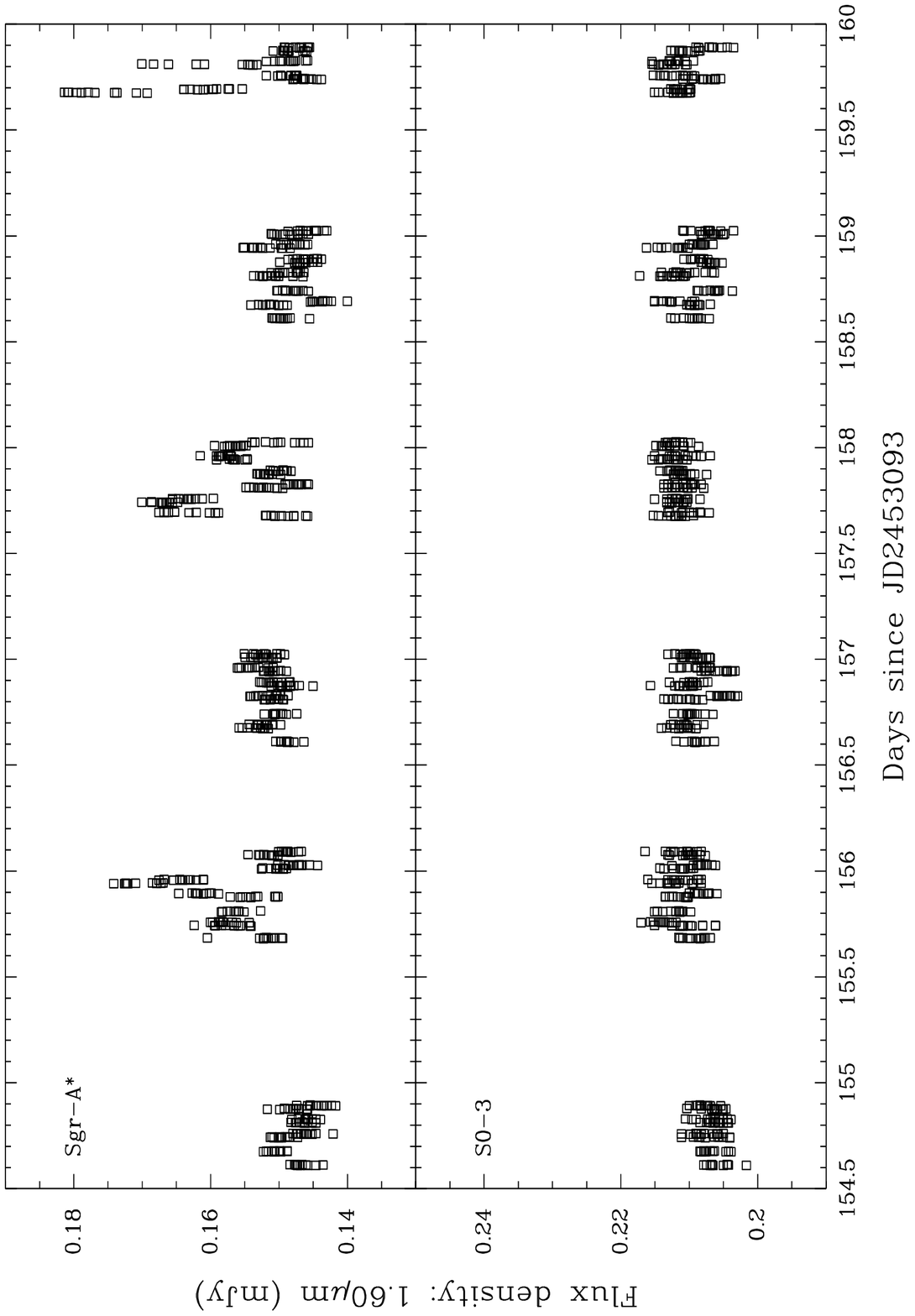}{3in}{-90}{50.0}{50.0}{-200}{280}
\plotfiddle{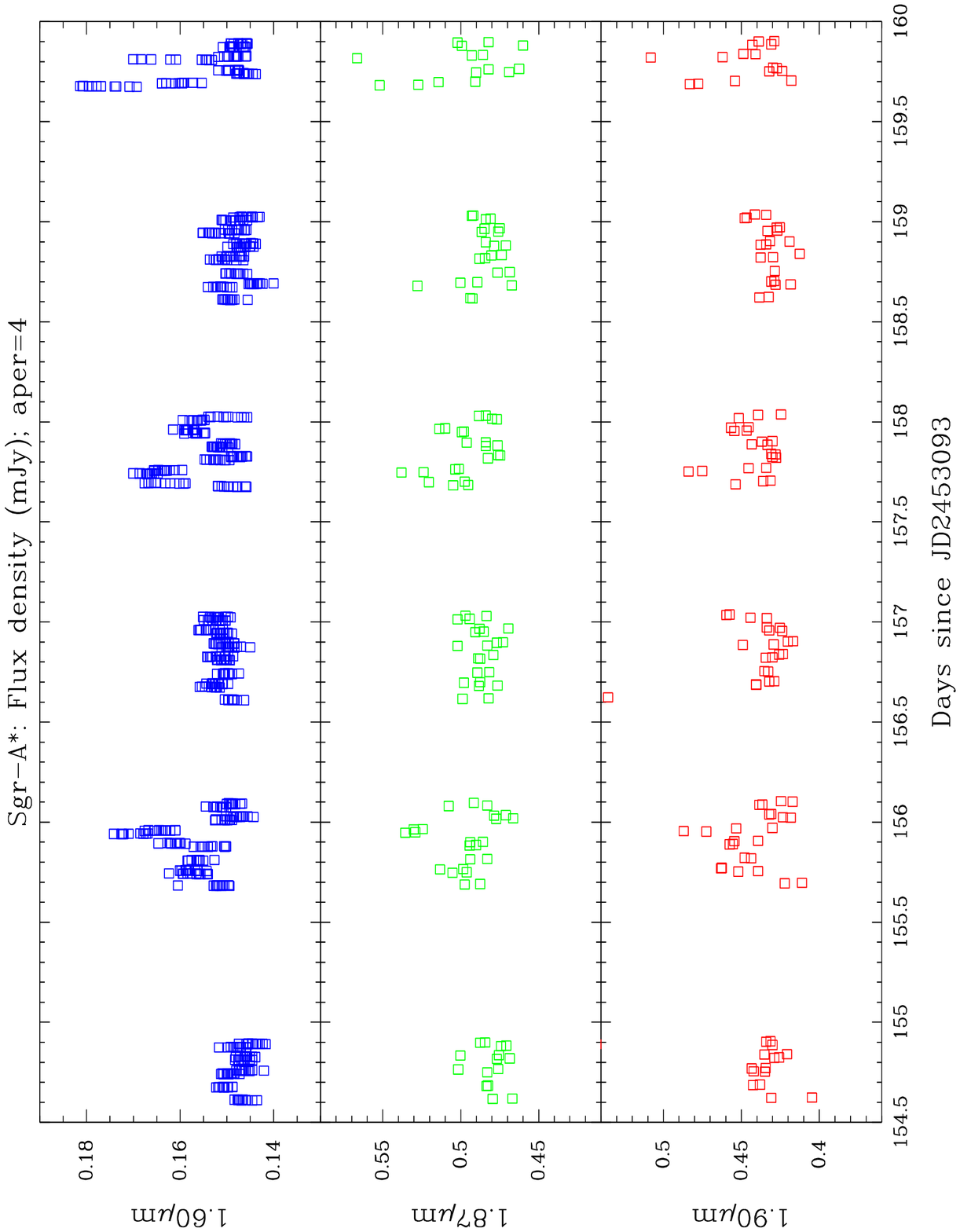}{3in}{-90}{45.0}{45.0}{-200}{250}
\caption{(a) The light curves of Sgr~A* and star S0-3 in the NICMOS
F160W band, using apertures of four pixels in diameter.
The data for Sgr~A* clearly show flare activity, while the signal
from S0-3 is quite constant (top panel).
(b) Near-IR
light curves of Sgr~A* using four-pixel diameter apertures are 
shown in six observing windows. Flare activity has been 
detected in three windows in all three bands. The fluxes
shown are the observed fluxes in each band, uncorrected for
reddening. Measurement uncertainties are $\sim$0.002 mJy in the
1.60 $\mu$m band and $\sim$0.005 mJy in the 1.87 and 1.90 $\mu$m
bands (bottom panel).}
\end{figure}
   
\begin{figure}
\plotfiddle{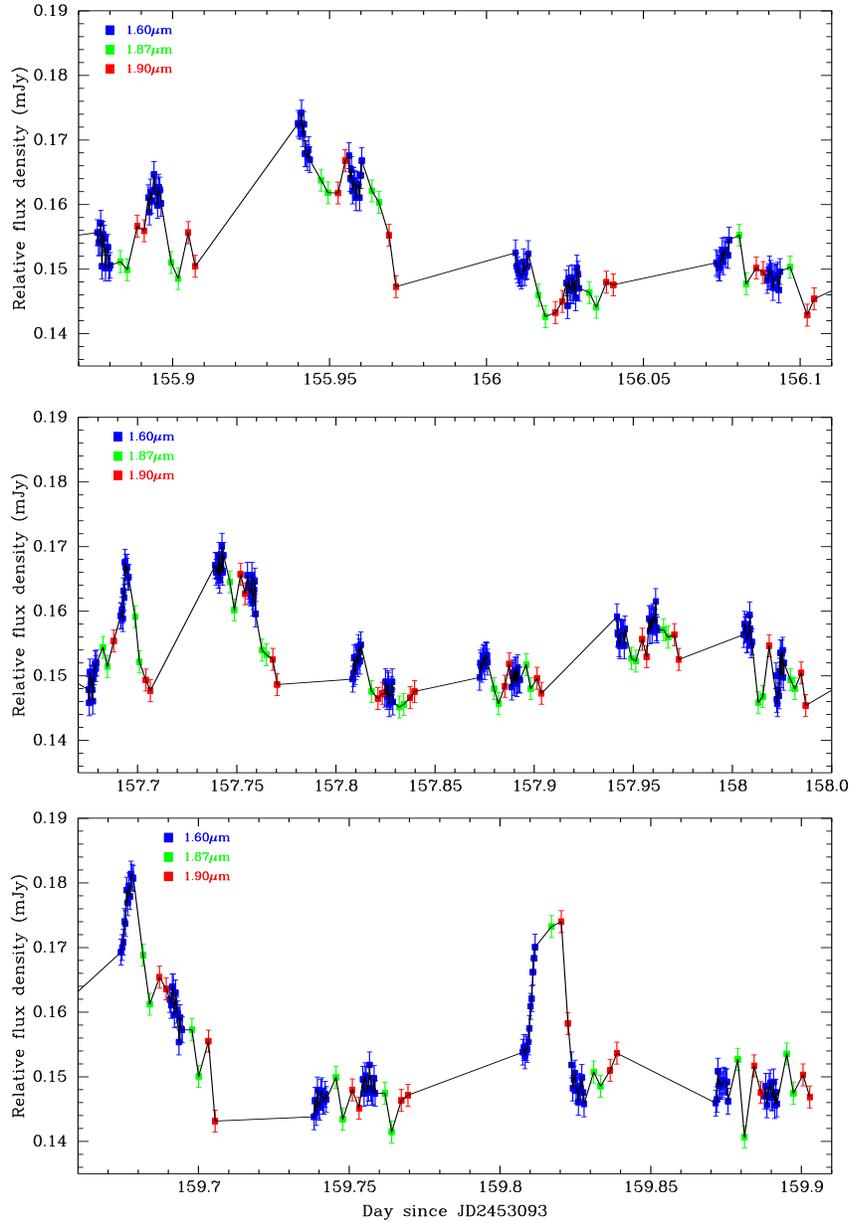}{6.5in}{0}{70.0}{70.0}{-160}{-80}
\caption{(a) Light curves of Sgr~A* in three different 
observing 
time windows during which near-IR flare activity took place. 
The blue, green, and red points represent the 1.60, 1.87, and 1.90 
$\mu$m bands, respectively. 
The broadband 1.60 $\mu$m data points are sampled every 32 sec, 
whereas the narrow-band 1.87 and 1.90 $\mu$m data 
are averages of six 32 sec data points, in order to get similar S/N 
ratio in the three bands. (b) 
This is a plot showing the simultaneous 2-Gaussian fit to the histogram of 
both the
noise and the flare emission at 1.6$\mu$m. 
The dotted lines are the 
individual
Gaussians, while the thick dashed line is the sum of the two. }
\end{figure}

\begin{figure}
\setcounter{figure}{2}
\plotone{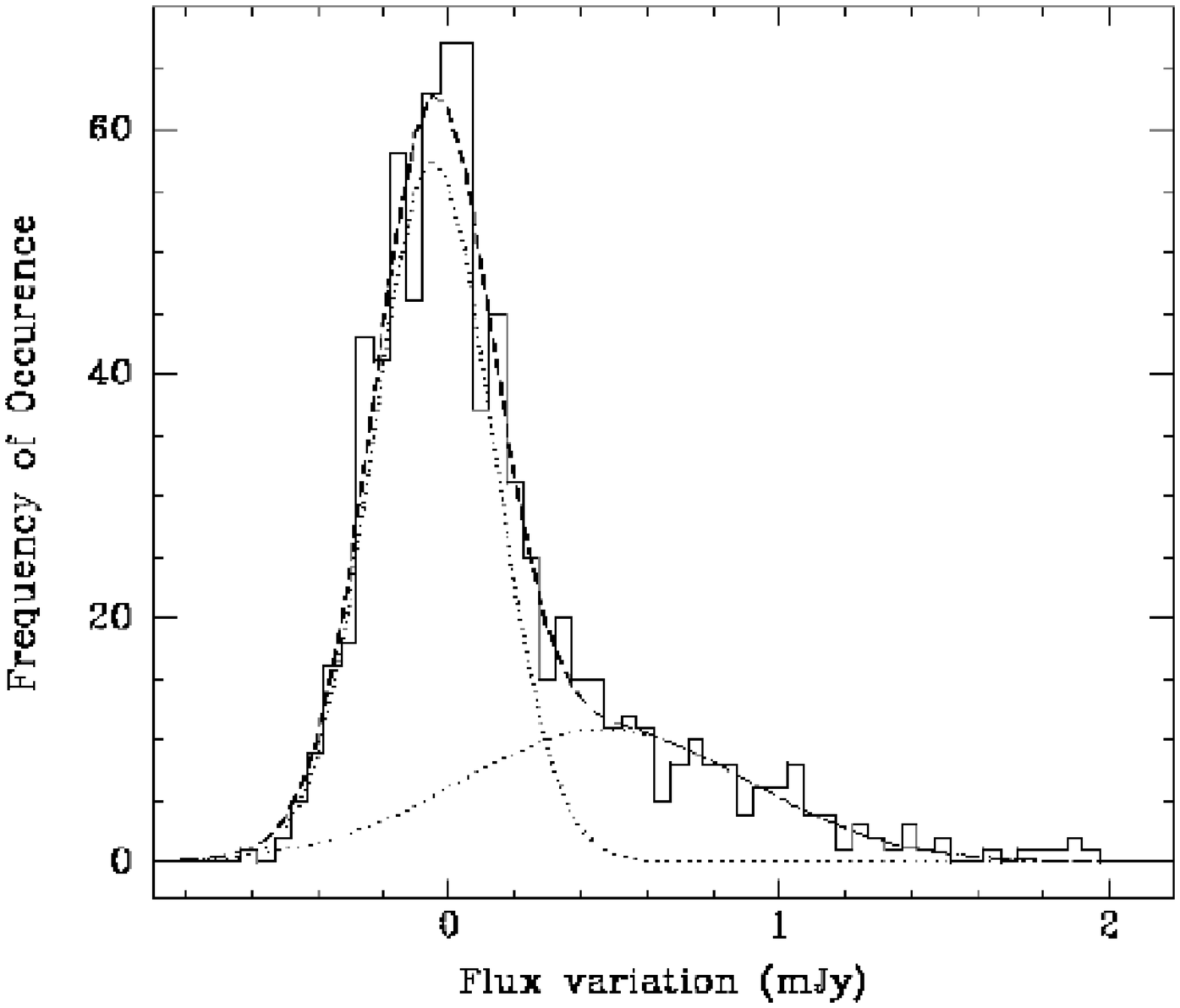}
\caption{}
\end{figure}

\begin{figure}
\plotfiddle{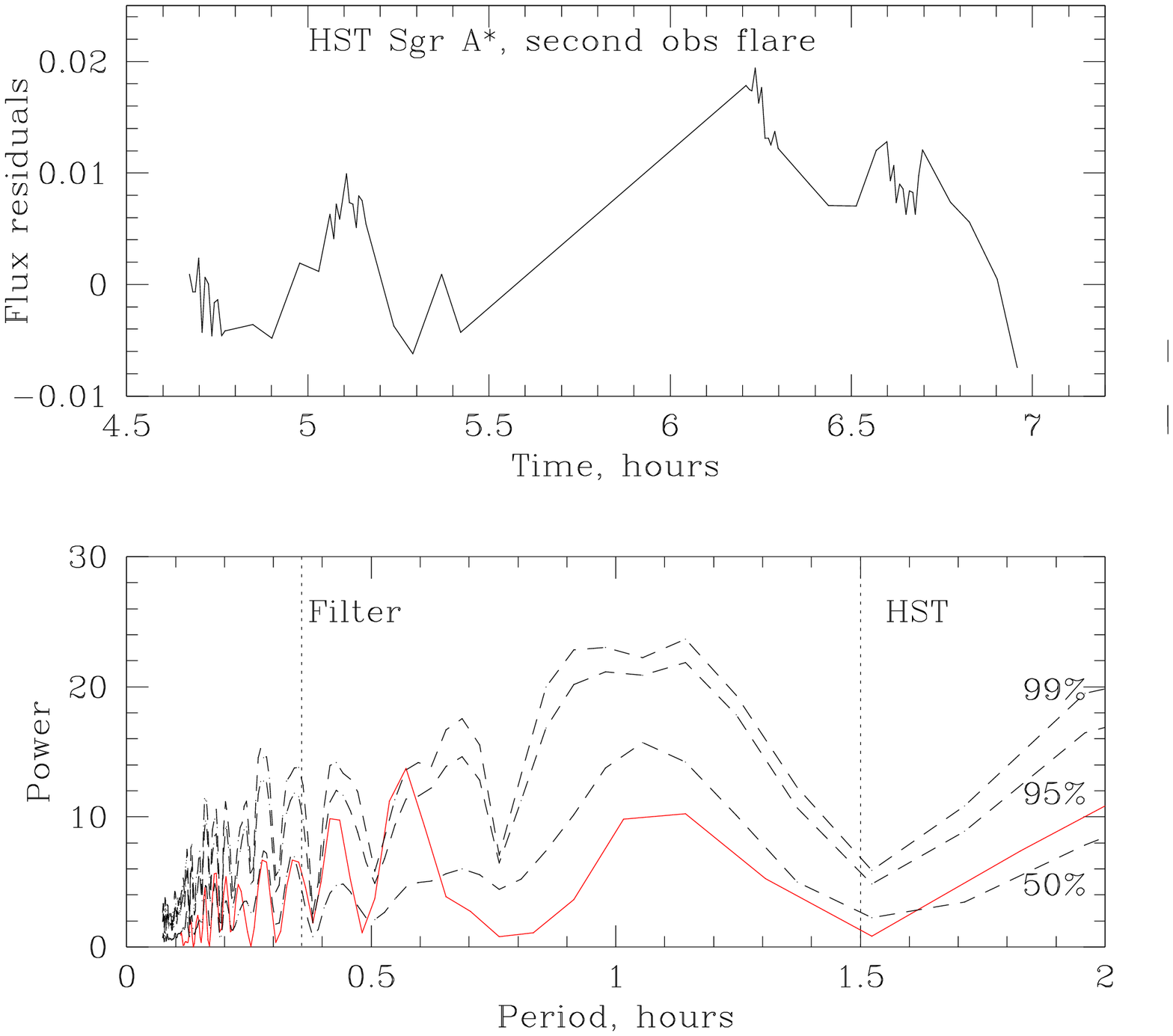}{3.7in}{0}{70}{60}{-238}{-110}
\plotfiddle{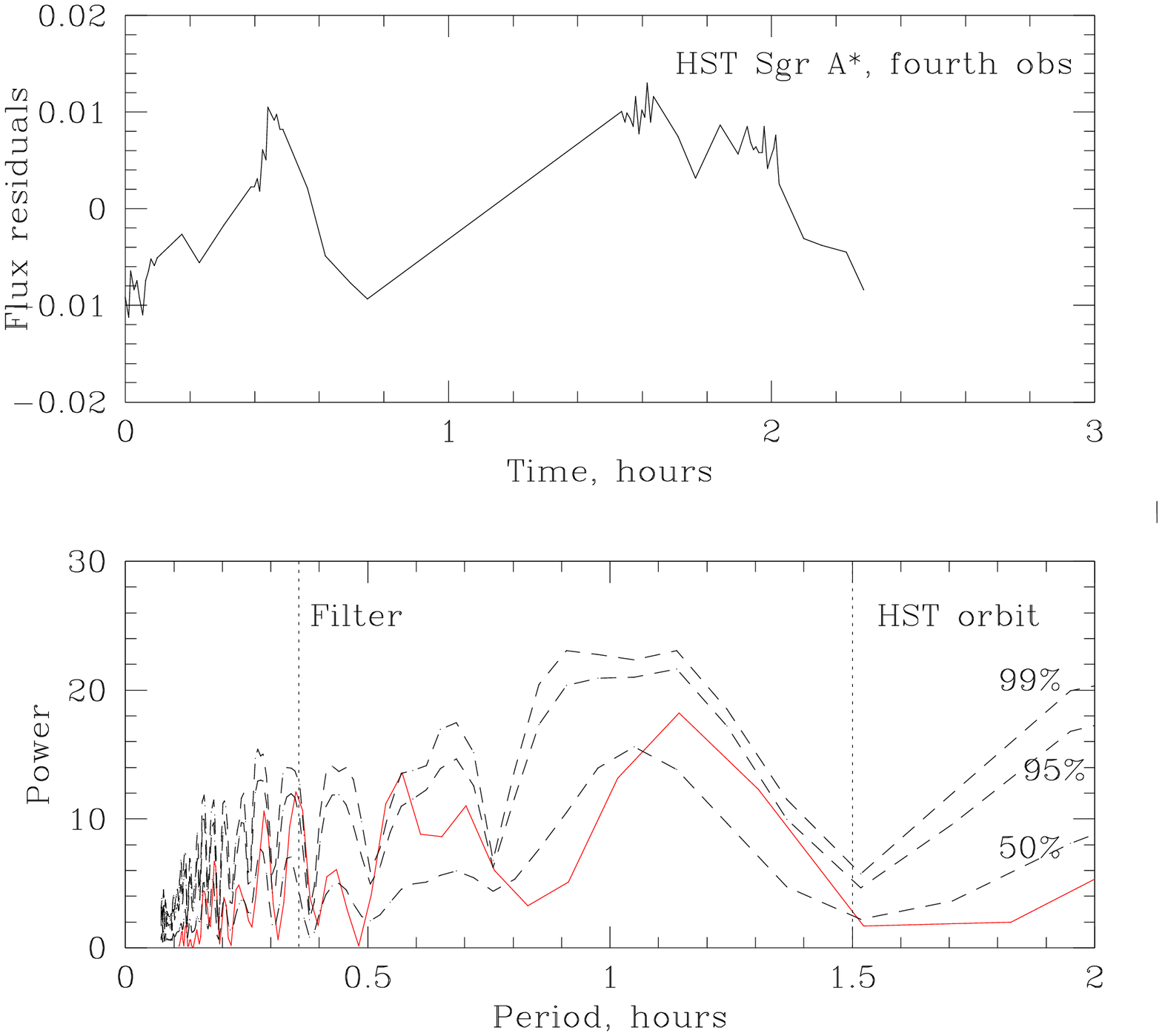}{3.7in}{0}{70}{60}{-238}{-110}
\clearpage
\caption{(a, top) The top and bottom boxes of the top panel 
 show the light curves 
and the corresponding power spectra of the residual flux of Sgr 
A* during  the
2nd  observing time window when flare activity was 
detected. The dashed lines 
show the significance of the power spectrum at  50, 95 and 99\% 
confidence levels. (b, bottom) Similar to (a)  except that the light curve and 
the power spectrum of the fourth observing time window are  shown. }
\end{figure}

\begin{figure}
\plotfiddle{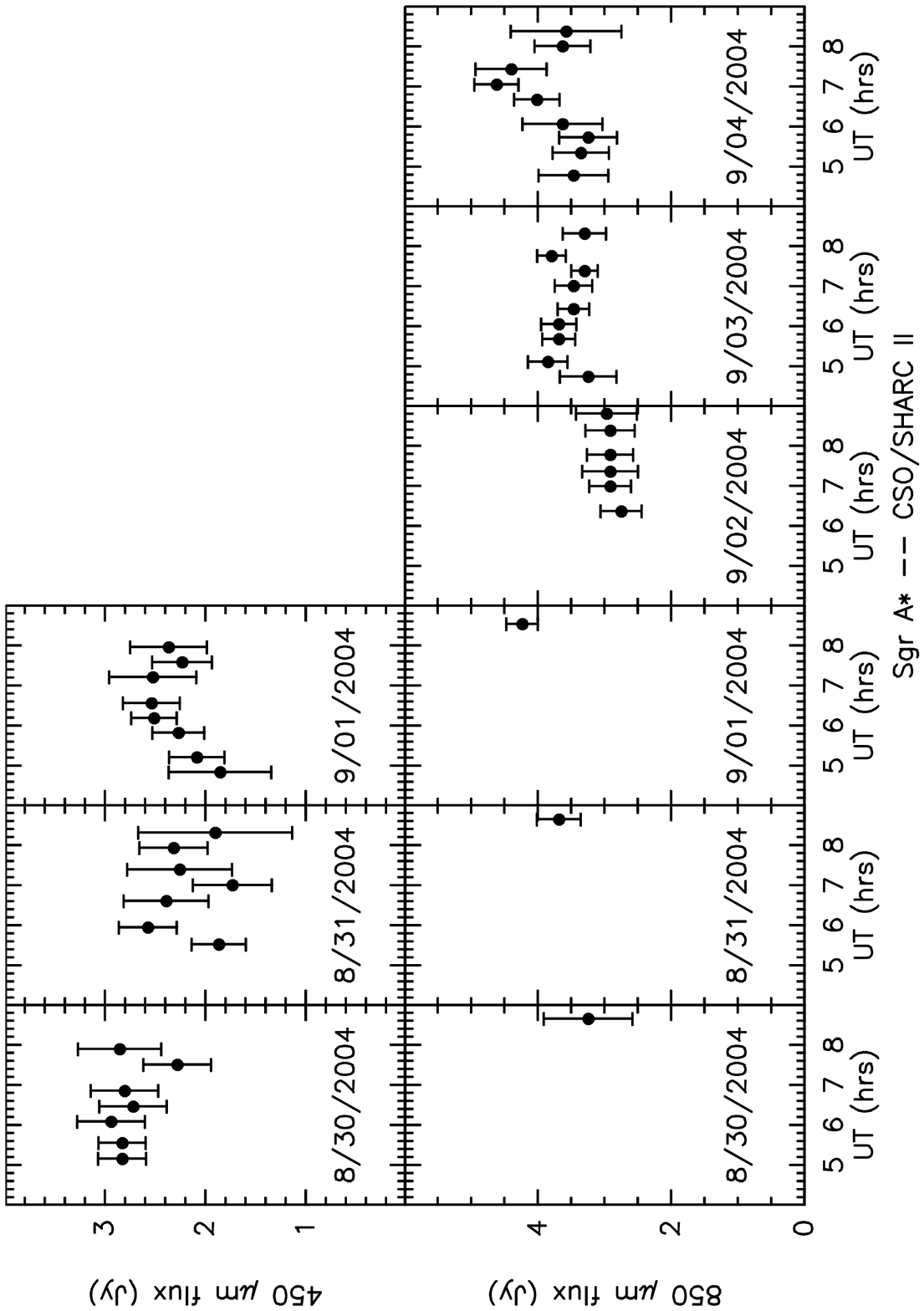}{3.7in}{-90}{70}{65}{-238}{330}
\plotfiddle{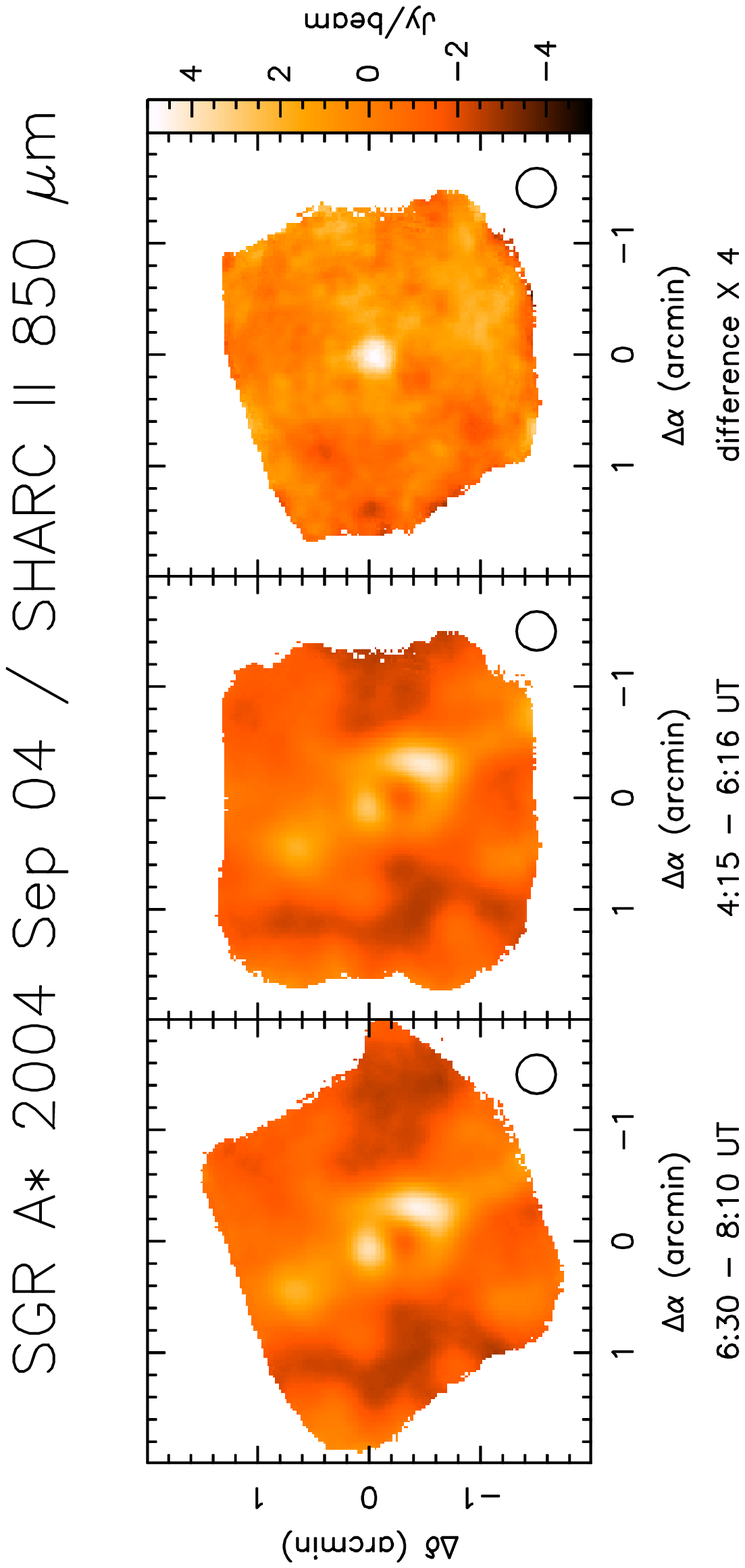}{3.7in}{-90}{74}{65}{-248}{360}
\clearpage
\caption{(a) The light curves of Sgr A* at 850 $\mu$m with 
1-$\sigma$ error bars during the second 
observing 
campaign with CSO. 
The integration time for each individual data point is 
20min (top panel). (b) The maps of  Sgr A* and the 
orbiting circumnuclear 
rings at 850$\mu$m in two different times during  the flare. 
The subtracted 
map to the 
right shows Sgr A* is the only significant source that was 
variable.      
}
\end{figure}

\begin{figure}
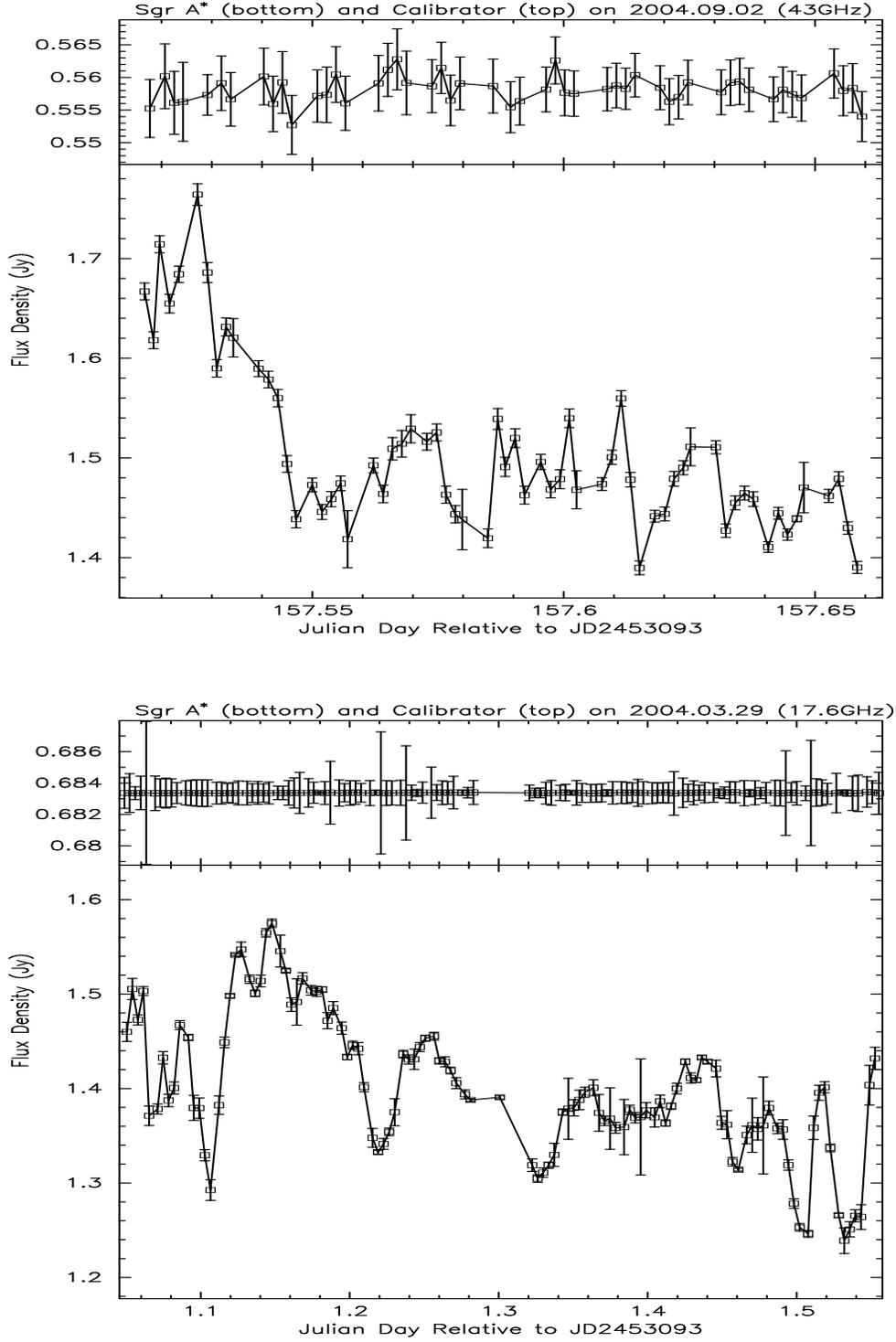

\plotfiddle{f6a.ps}{3.7in}{0}{70}{40}{-238}{0}
\plotfiddle{f6b.ps}{3.7in}{0}{70}{40}{-238}{0}
\caption{The complex gain of the calibrators 
as a function of time and the corresponding light curves of
Sgr A* based on VLA and ATCA observations at 7mm and 1.2cm are shown 
in the top (a) and bottom panels (b), respectively.  }
\end{figure}

\begin{figure}
\plotone{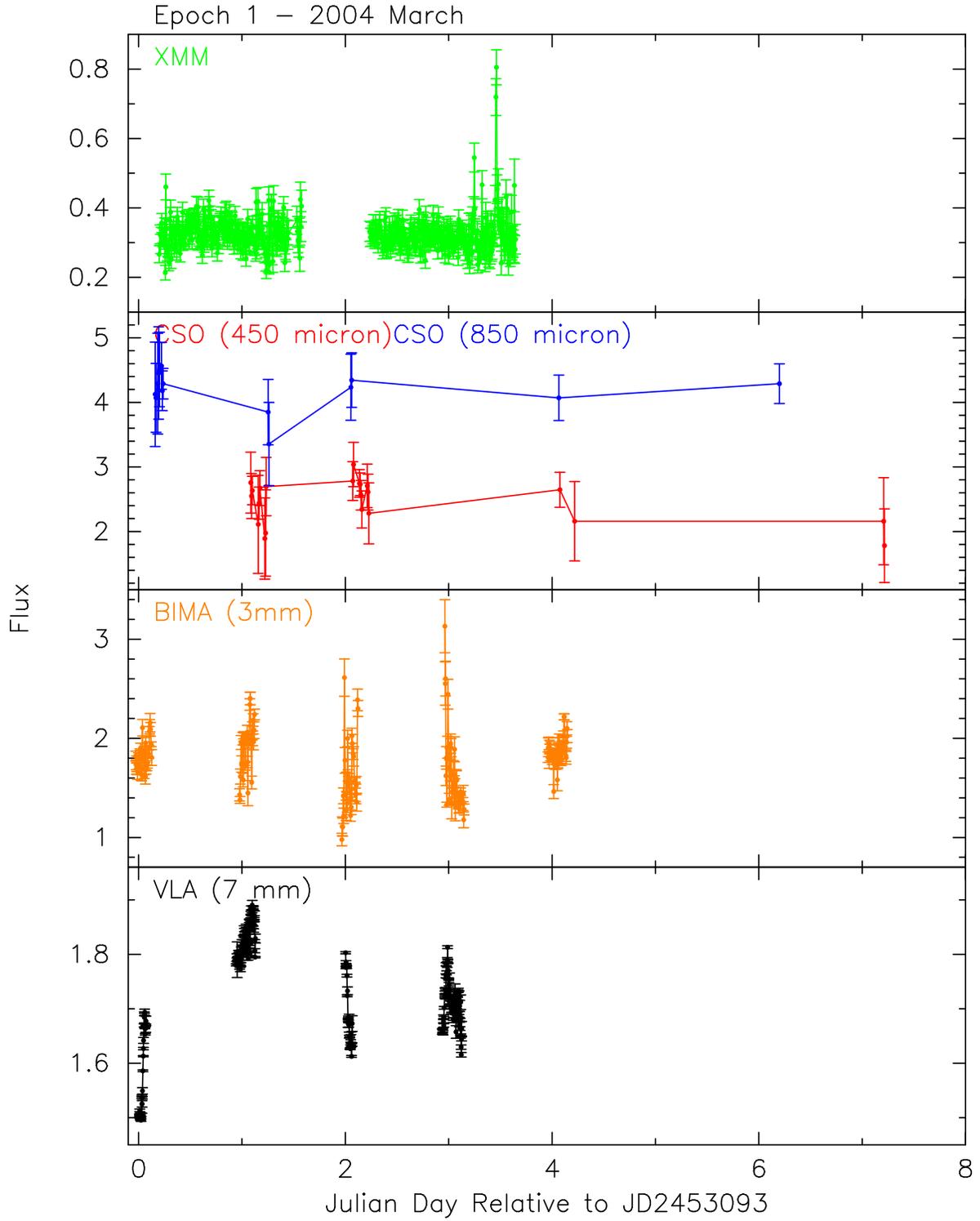}
\caption{The simultaneous X-ray, submillimeter, millimeter and radio 
emission from Sgr A* based on the first epoch of observations using 
XMM, CSO, BIMA and VLA. }
\end{figure}

\begin{figure}
\plotone{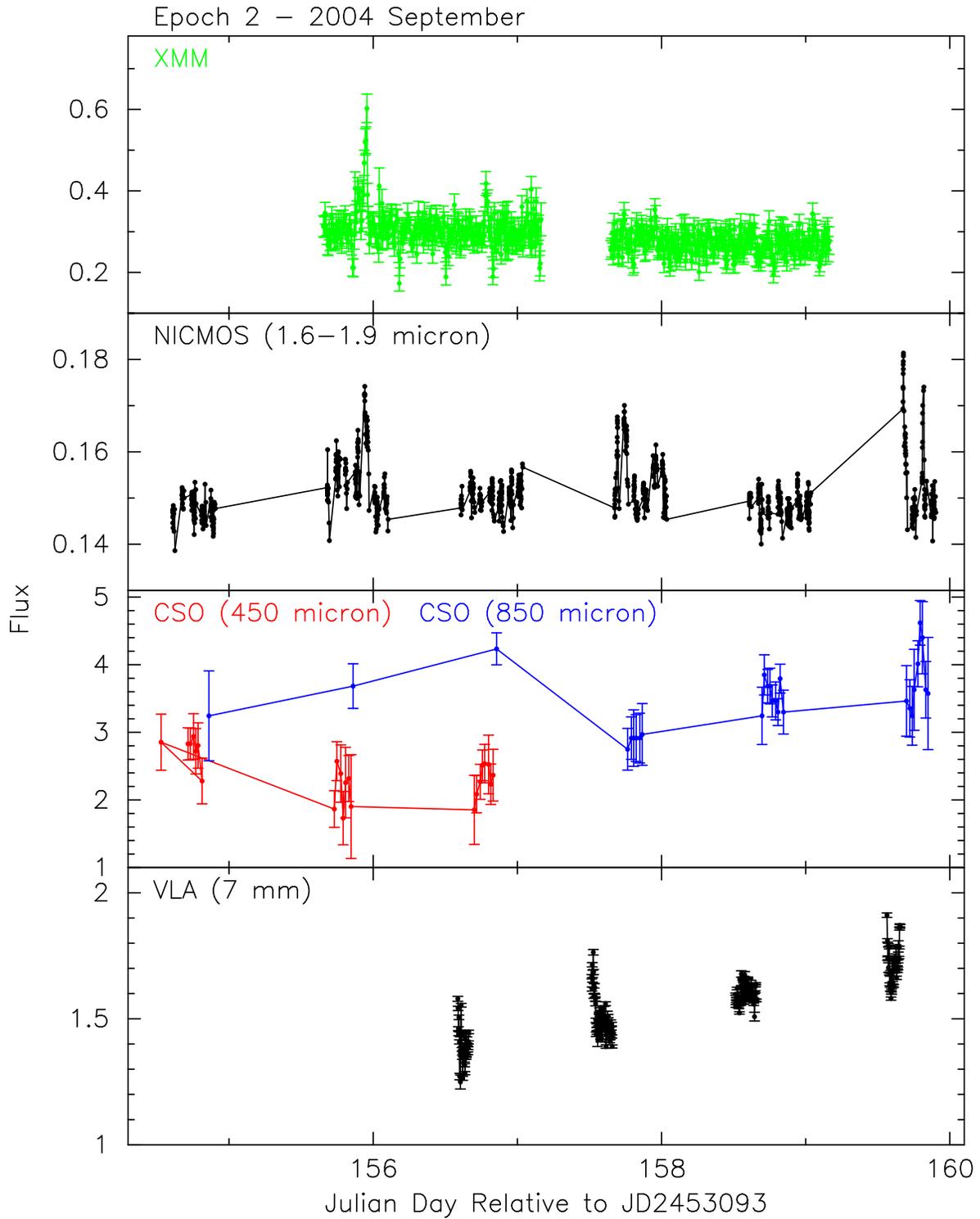}
\caption{The simultaneous X-ray, near-IR, submillimeter, and radio 
emission from Sgr A* based on the second  epoch of observations using 
XMM, HST, CSO, and VLA. The 8h periodic dips  
detected in the X-ray light 
curve are due to the eclipses of the transient, as described in Porquet 
et al. 2005).}
\end{figure}

\begin{figure}
\plotfiddle{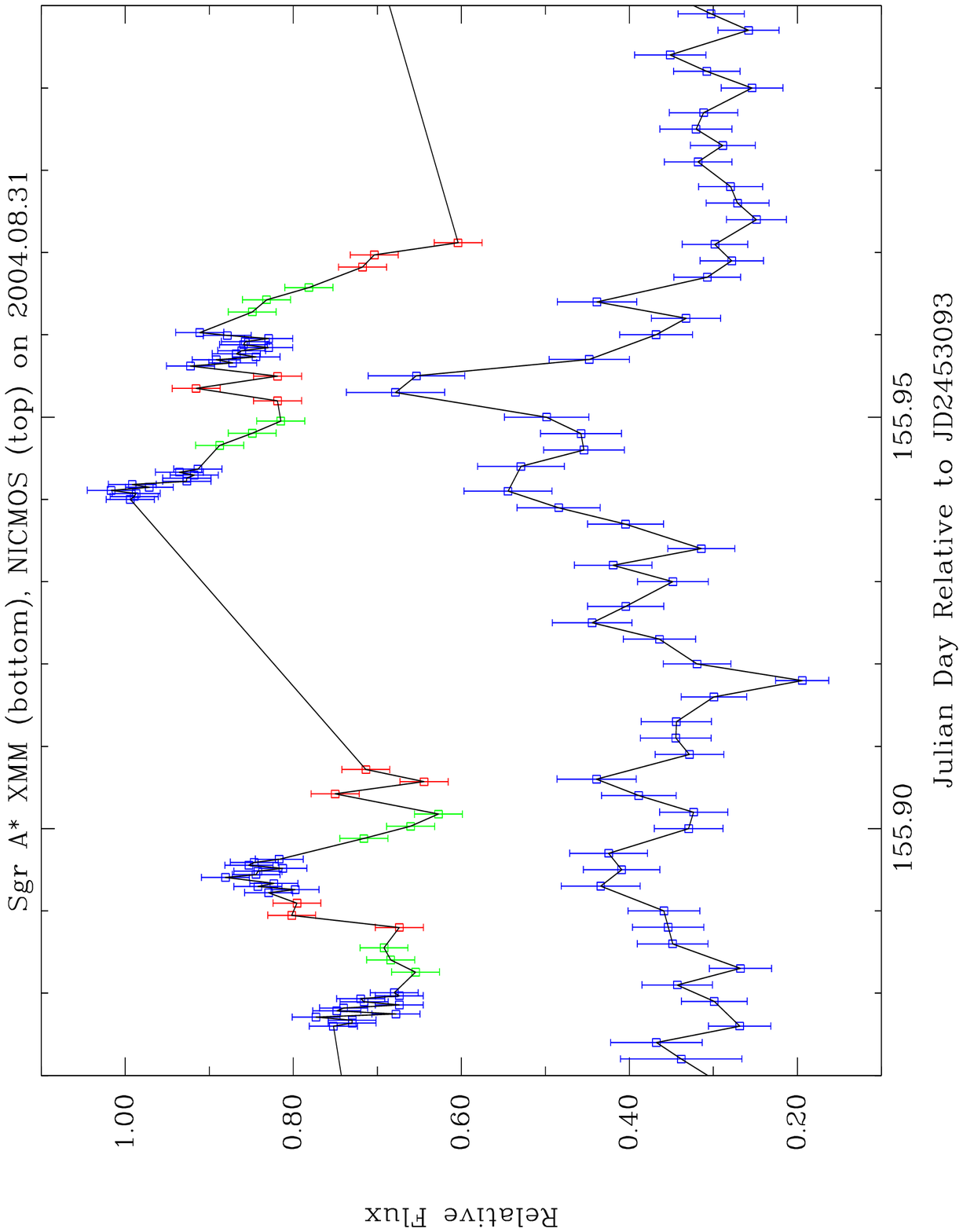}{6.5in}{-90}{70}{70}{-300}{500}
\clearpage
\caption{The simultaneous near-IR (top plot) and X-ray (bottom plot) 
light curves 
during which flares were  
detected in the 
second 
epoch of the observing campaign. The near-IR flares are detected in the 
2nd observing window,  as shown in Figure 3a. 
The vertical axis shows 
the observed flux density in  near-IR wavelengths whereas the observed 
count  rate  
in X-ray wavelengths. 
The F187N (green color) and F190N (red color) filter data are sampled at 
128 sec intervals
whereas the F160W (blue color) are sampled every 32 sec. The X-ray data 
have 200-second  time resolution.  }
\end{figure}

\begin{figure}
\plotone{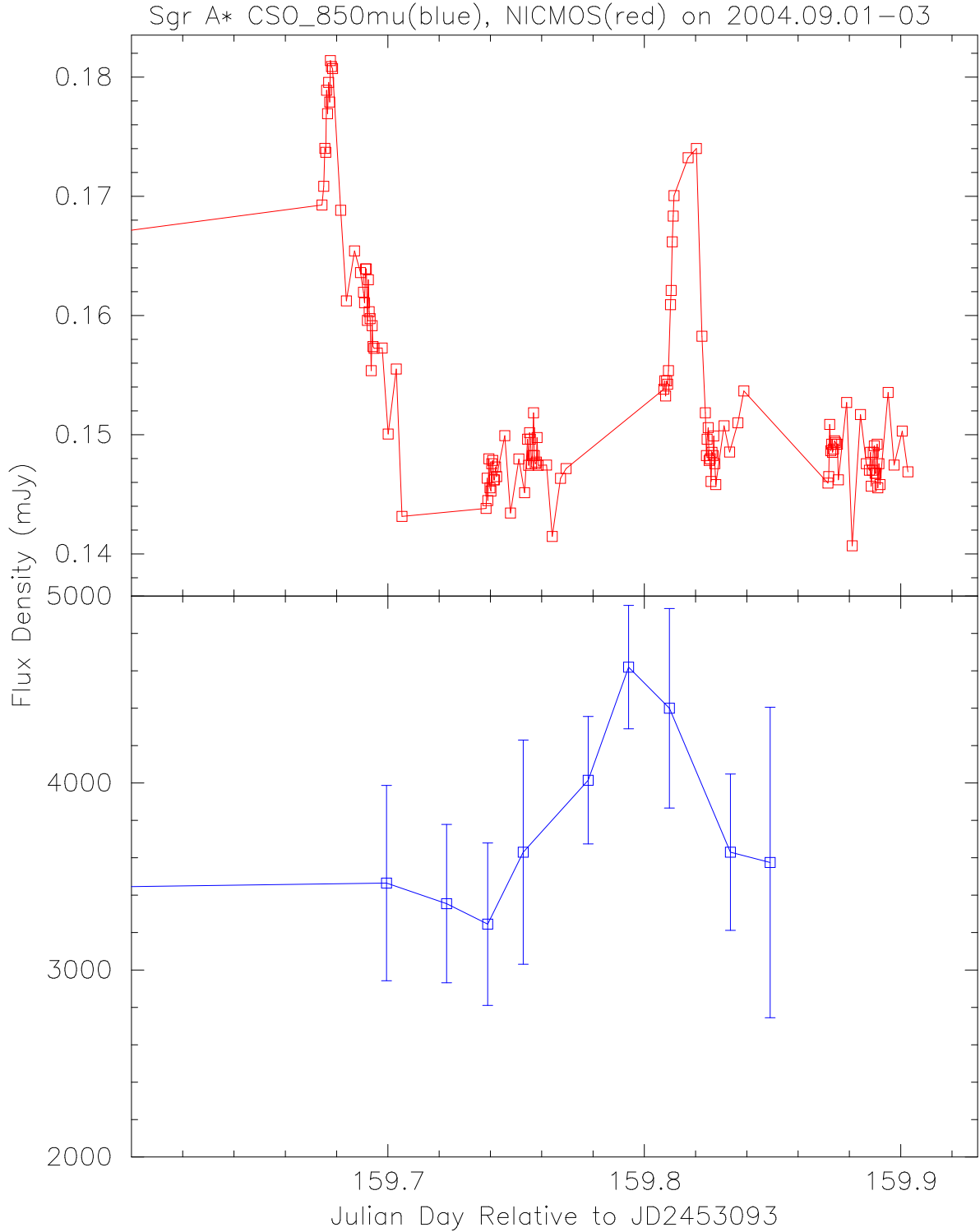}
\caption{The simultaneous near-IR 
and 850$\mu$m  light curves showing  flares detected in the 
second 
epoch of the 
observing campaign. The near-IR flares correspond to the 
6th  observing window,  as shown in Figure 3. There are no  X-ray observations 
during the period when these flares took place.  }
\end{figure}

\begin{figure}
\plottwo{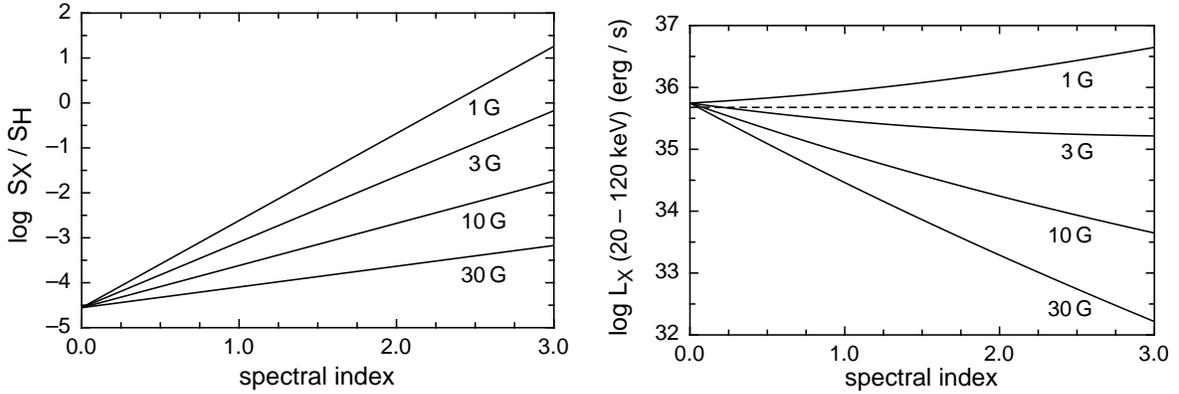}{f11b.epsi}
\caption{(a)[left] For a given magnetic field, a run of the 
log 
of the ratio of the fluxes at 1 keV and 1.6$\mu$m 
from Sgr A* is drawn as  a 
function of the near-IR spectral 
index, for the scattering  of submillimeter 
emission via inverse 
Compton.  
(b)[right] Similar to Figure (a) except that the vertical axis 
shows the 
log of the luminosity between 2 - 120 keV. The dashed line shows the 
measured flux from the inner 13$'$ of the Galactic center.}
\end{figure}

\begin{figure}
\plotfiddle{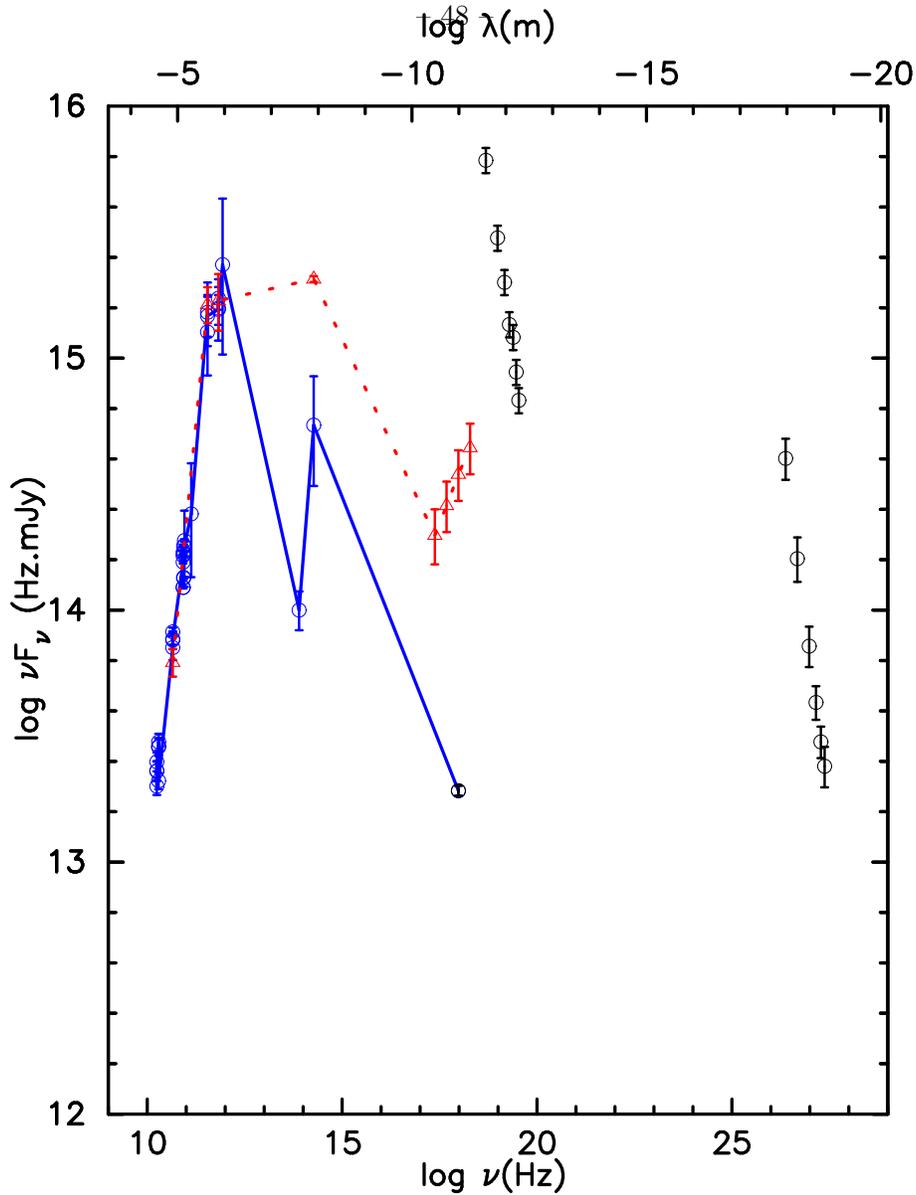}{5.5in}{0}{85}{85}{-250}{-110}
\caption{The spectrum of Sgr A* based on all the measurements taken during 
the first epoch of observations is shown in blue color. This clearly shows 
that the 
quiescent flux 
of Sgr A* peaks in submillimeter wavelength between 0.45 and 2mm, probably 
near 0.1 mm. The spectrum of Sgr A* is based on 
measurements at 
7mm, 450$\mu$m, 850$\mu$m, 3.6$\mu$m,  
1.6$\mu$m and X-rays at 1 keV. The quiescent X-ray and near-IR fluxes are  
taken from 
Baganoff et al. (2001), Genzel et al. (2003) and Ghez et al. (2005), 
though  there may not be a true quiescent flux in near-IR wavelengths. 
All the measurements shown in red color  are associated with  flaring activity
including the X-ray flare with a  spectral index of 0.6  detected by XMM (Belanger et al. 
2005a).   
The hard $\gamma$-ray 
emission from the Galactic center at 1 TeV based on HESS observations 
(Aharonian et al. 2004)  and 
the high  fluxes from 
INTEGRAL observations are not
connected 
and are only shown by themselves in black color with the corresponding spectral index of 
2.2 and 2.06, respectively. The mean 
flux and the 1-$\sigma$ 
error are 
determined from data 
taken closest to the simultaneous near-IR/X-ray flare. }
\end{figure}

\end{document}